\documentclass[pdftex,twocolumn,epjc3,envcountsect]{svjour3}          

\RequirePackage[T1]{fontenc}
\usepackage[american]{babel}
\usepackage[normalem]{ulem}

\smartqed  

\RequirePackage{graphicx}
\RequirePackage{mathptmx}      
\RequirePackage{flushend}
\RequirePackage[numbers,sort&compress]{natbib}
\RequirePackage[colorlinks,citecolor=blue,urlcolor=blue,linkcolor=blue]{hyperref}

\usepackage{amsmath}
\usepackage{amssymb}
\usepackage{subfigure}
\usepackage{tabularx}
\usepackage{booktabs}
\usepackage{dcolumn}
\usepackage{bm}
\usepackage{xcolor}
\usepackage{epstopdf}
\usepackage{dsfont}
\usepackage{comment}
\usepackage{xstring}

\usepackage{microtype} 

\usepackage{cleveref}
\Crefname{section}{Section}{Sections}
\Crefname{figure}{Figure}{Figures}
\Crefname{table}{Table}{Tables}
\Crefname{appendix}{Appendix}{Appendices}
\Crefname{equation}{Eq.}{Eqs.}
\newcommand{\Onlinecite}[1]{%
    \IfSubStr{#1}{,}{Refs}{Ref}.~\cite{#1}%
}

\DeclareMathOperator{\tr}{\rm tr\,}

\journalname{Eur. Phys. J. C}

\begin{document}

\title{Unveiling confinement in pure gauge SU(3):
flux tubes, fields,  and magnetic currents}

\author{M. Baker\thanksref{e1,addr1}
\and
V. Chelnokov\thanksref{e2,addr2,addr6}
\and
L. Cosmai\thanksref{e3,addr3}
\and
F. Cuteri\thanksref{e4,addr2}
\and
A. Papa\thanksref{e5,addr4,addr5}
}

\institute{Department of Physics, University of Washington, WA 98105 Seattle, USA\label{addr1}
\and
Institut f\"ur Theoretische Physik, Goethe Universit\"at, 60438 Frankfurt am Main, Germany\label{addr2}
\and
INFN - Sezione di Bari, I-70126 Bari, Italy\label{addr3}
\and
Dipartimento di Fisica, Universit\`a della Calabria, I-87036 Arcavacata di Rende, Cosenza, Italy\label{addr4}
\and
INFN - Gruppo collegato di Cosenza, I-87036 Arcavacata di Rende, Cosenza, Italy\label{addr5}
\and
{\it{on leave of absence from}} Bogolyubov Institute for Theoretical Physics of the National Academy of Sciences of Ukraine\label{addr6}
}

\thankstext{e1}{e-mail: mbaker4@uw.edu}
\thankstext{e2}{e-mail: chelnokov@itp.uni-frankfurt.de}
\thankstext{e3}{e-mail: leonardo.cosmai@ba.infn.it}
\thankstext{e4}{e-mail: cuteri@itp.uni-frankfurt.de}
\thankstext{e5}{e-mail: alessandro.papa@fis.unical.it}

\date{Received: date / Accepted: date}

\maketitle

\begin{abstract}
 A characteristic signature of quark confinement is the concentration of the chromoelectric field between a static quark-antiquark pair in a flux tube. However, the structure of this flux tube, and hence of the confining force, has not been completely understood. Here we perform new lattice measurements of field distributions on smeared Monte Carlo ensembles in SU(3) gauge theory. On the basis of these simulations we demonstrate that the confining force can be understood using the analogy with the basic principles of electromagnetism as elucidated by Maxwell.
 We derive a 
 chromomagnetic Lorentz force density coupling the chromoelectric field to chromomagnetic currents and integrate this force density over the flux tube interior to obtain a Maxwell-like force that squeezes the flux tube in the transverse direction. We show that the strength of this transverse confining force is equal to the value of the string tension calculated numerically from  the chromoelectric field on the midplane between the quarks, verifying the consistency of these two complementary pictures of confinement.
\end{abstract}

\section{Introduction}

The confinement of quarks and gluons inside hadrons remains an open problem of Quantum Chromodynamics (QCD). A theoretical explanation of this phenomenon is still missing and our current understanding is based on models of the QCD vacuum (for a review, see 
\Onlinecite{greensite2011introduction,Diakonov:2009jq}) and Monte Carlo numerical simulations
of QCD on a space-time lattice.

A great deal of numerical evidence shows  that a static quark and antiquark interact via a confining linear potential for distances equal to or larger than about 0.5~fm. 
This linear potential is almost entirely due to the electric 
\footnote{Here and further we refer to the chromoelectric and chromomagnetic fields as just 
 ``electric'' and ``magnetic''.}
field, which is mostly longitudinal, {\em i.e.},  oriented along the line connecting the static quark and
antiquark~\cite{Bander:1980mu,Greensite:2003bk,Ripka:2005cr,Simonov:2018cbk}.

Many numerical studies in SU(2) and SU(3) Yang-Mills theories~\cite{Fukugita:1983du,Kiskis:1984ru,Flower:1985gs,Wosiek:1987kx,DiGiacomo:1989yp,DiGiacomo:1990hc,Cea:1992sd,Matsubara:1993nq,Cea:1994ed,Cea:1995zt,Bali:1994de,Green:1996be,Skala:1996ar,Haymaker:2005py,D'Alessandro:2006ug,Cardaci:2010tb,Cea:2012qw,Cea:2013oba,Cea:2014uja,Cea:2014hma,Cardoso:2013lla,Caselle:2014eka,Cea:2015wjd,Cea:2017ocq,Shuryak:2018ytg,Bonati:2018uwh,Shibata:2019bke} have characterized the shape of the electric field in the transverse plane at the midpoint of the line connecting the static quark and antiquark. More recently, numerical investigations have extended their reach, achieving a detailed description of all components of the color fields around static sources~\cite{Baker:2018mhw,Baker:2019gsi}, as well as the spatial
distribution of the stress energy-momentum tensor~\cite{Yanagihara:2018qqg,Yanagihara:2019foh,Baker:2019gsi} and the flux densities for hybrid static potentials~\cite{Bikudo:2018,Mueller:2019mkh}.      
This growing numerical phenomenology about color fields near static sources could give new hints in quest for the mechanism of confinement.

The present paper extends the results of \Onlinecite{Baker:2018mhw,Baker:2019gsi}. The main message
in those papers was that the electric field can be viewed as the superposition of a
`perturbative' part, which is the sole contributor to  components  transverse 
 the quark-antiquark axis, and a `non-perturbative',  longitudinal part, all components of the magnetic field being negligibly small everywhere.
Here we focus instead on the distribution of electric and magnetic sources and currents, as they can be inferred from the shape of color fields. In particular, we present for the first time evidence of the solenoidal magnetic current responsible for the longitudinal electric field and study its behavior toward the continuum limit, leading to new understanding of confinement.

The organization of the paper is as follows: \Cref{sec:correlator,sec:separation,sec:Maxwell,sec:magnetic} are devoted to the theoretical background, \Cref{sec:setup} describes our numerical setup, \Cref{sec:numerical} contains our new results, and conclusions are drawn in \Cref{sec:conclusion}.

\section {Connected correlator and the field strength tensor}
\label{sec:correlator}

In previous papers~\cite{Baker:2018mhw,Baker:2019gsi},  using lattice simulations of SU(3) pure gauge theory we have measured the spatial distributions of the color fields induced by a quark-antiquark pair separated by a range of distances $d$ ranging from 0.37~fm to 1.25~fm.

\begin {figure}[htb]
\centering
\includegraphics[width=0.7\linewidth,clip]{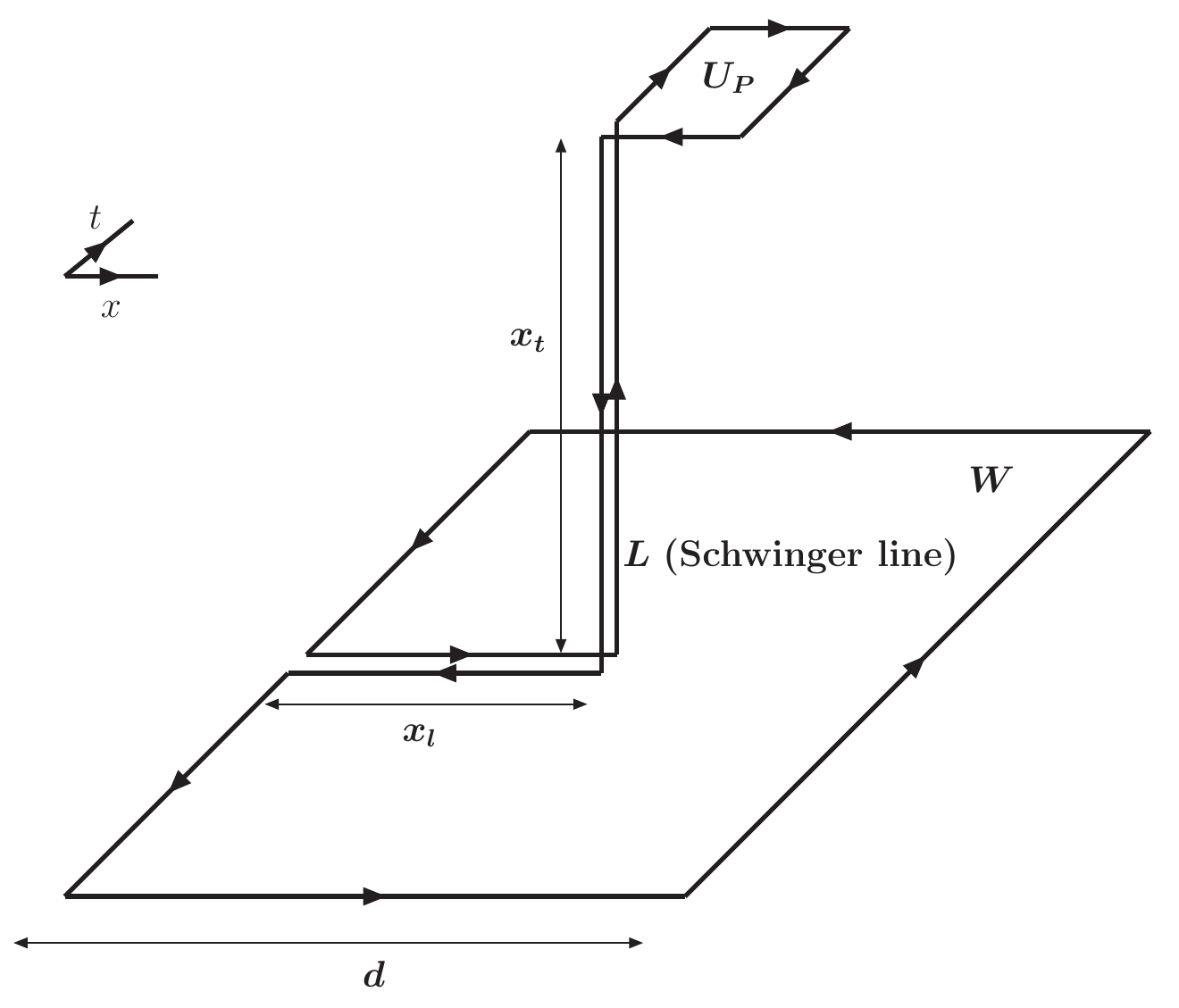}
\includegraphics[width=0.25\textwidth,clip]{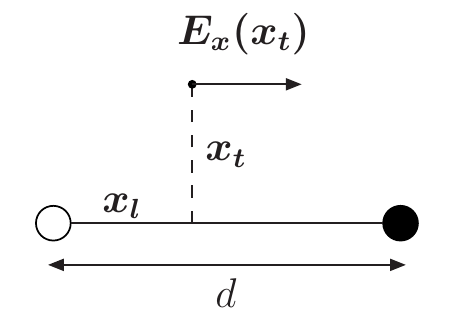}
\caption{The connected correlator between the plaquette $U_{P}$ and the Wilson loop (subtraction in $\rho_{W,\,\mu\nu}^{\rm conn}$ not explicitly drawn).
The longitudinal electric field $E_x(x_t)$ relative to the position of the static sources (represented by the white and black circles), for a given value of the transverse distance $x_t$.  ($x_l = x,~x_t = y$.)}                        \label{fig:op_W}
\end{figure}

These distributions were obtained from lattice measurements of the connected correlation function $\rho^{\rm conn}_{W, \mu \nu}$~\cite{DiGiacomo:1989yp} of a plaquette $U_P = U_{\mu \nu} (x)$ in the ${\mu \nu}$ plane, and a Wilson loop $W$ in the $\hat 4\hat 1$  plane (see \Cref{fig:op_W}),
\begin{equation}
    \rho^{\rm conn}_{W, \mu \nu} = \frac {\langle\tr (WLU_PL^*)\rangle}{\langle\tr(W)\rangle} - \frac{1}{N} \frac {\langle\tr (U_P) \tr (W)\rangle}{\langle\tr(W)\rangle}\;,
    \label{connected1}
\end{equation}
$N=3$ being the number of QCD colors.

The correlator $\rho^{\rm conn}_{W, \mu \nu}$ provides a lattice definition of a gauge-invariant field strength tensor $\langle F_{\mu \nu}\rangle_{q \bar{q}} \equiv F_{\mu \nu}$ carrying a unit of octet charge, while possessing the space-time symmetry properties of the Maxwell field tensor of electrodynamics. 
\begin{equation}
\rho^{\rm conn}_{W, \mu \nu} \equiv~~ a^2 g\langle F_{\mu \nu}\rangle_{q \bar{q}} ~~ \equiv~~ a^2 g ~F_{\mu \nu}\;.
\label{connected2}
\end{equation}

When the plaquette $U_P$ lies in the $\hat 4 \hat 1$ plane, the measured $\hat 4 \hat 1$ component of the field tensor determines $E_x$, the component of the electric field along the $q\bar{q}$ axis $ E_x = F_{41}$; {\em i.e.}, the longitudinal component of the electric field at the position corresponding to the center of the plaquette. 

When $U_P$ is in the $\hat 4 \hat 2$ plane, $F_{42} = E_y$, a component of the electric field transverse to the $q\bar{q}$ axis, and
when $U_P$ is in the $\hat 2 \hat 3$ plane, $F_{23} = B_x$,  the  longitudinal component of the magnetic field.
In the numerical evaluations of \Onlinecite{Baker:2018mhw,Baker:2019gsi} all components $B_i $  of the magnetic field were equal to zero within statistical errors.

\section {Separation of the electric field into `perturbative' and `non-perturbative' components}
\label{sec:separation}

The transverse components $E_y$ of the simulated electric field $\vec{E}$ were fit (\Onlinecite{Baker:2019gsi}) to the transverse components of an effective `perturbative' Coulomb-like field $\vec{E}^{\rm C}$ satisfying the condition:
\begin{equation}
    \vec{\nabla} \times \vec{E}^{\rm C} =0 \;.
    \label{curl}
\end{equation}
Subtracting $\vec{E}^{\rm C}$ from the simulated field $\vec{E}$ gives the 
`non-perturbative' longitudinal field $\vec{E}^{\rm NP}$ pointing along the $q\bar{q}$ axis $\hat{e}_x$:
\begin{equation}
    \vec{E}^{\rm NP} = \vec{E} - \vec{E}^{\rm C}\;.
    \label{ENP} 
\end{equation}

Throughout this paper all the field derivatives ({\em e.g.}, $\vec{\nabla} \times \vec{E}$ and $\vec \nabla \cdot \vec{E}$) are calculated numerically on the lattice from the measured electric field $\vec{E} (x_l, x_t)$. See Fig (\ref{fig:op_W}).

\section {The `Maxwell' picture of the Yang-Mills flux tube}
\label{sec:Maxwell}

In our previous paper \cite{Baker:2019gsi} we showed that the string tension $\sigma$ can be obtained from lattice simulations of the connected correlator $\rho^{\rm conn}_{W, \mu \nu}$, given in~\Cref{connected2}, defining the Maxwell-like field $F_{\mu\nu}$, 
as the integral over the midplane  $x_l = d/2$  between the quarks of the longitudinal component $T_{xx}$ of the Maxwell stress tensor,
\begin{equation}
 T_{\alpha \beta} = F_{\alpha \lambda} F_{\beta \lambda} - \frac{1}{4} \delta_{\alpha \beta} F_{\mu \lambda} F_{\mu \lambda} \;,
 \label{stress}   
\end{equation}
constructed from the non-perturbative electric field $\vec{E}^{NP} (x_l, x_t)$: 
\begin {equation}
\sigma =  \int_{x_l = d/2}  \mathrm{d}^2 x_t T_{xx}^{NP}\;,
\label{integralT}
\end{equation}
where
\begin{equation}
T_{xx}^{NP}= (\vec{E}^{\rm NP}(x_l = d/2,  x_t))^2/2\;.
\label{maxwell}
\end{equation}
Independently, the string tension can be determined from the value of $\vec{E}^{\rm NP}$ at the position of the quark. 

In this paper we use the lattice simulations of $F_{\mu \nu}$  to evaluate the divergence of the
 Maxwell stress tensor  $T_{\alpha \beta}$ and obtain the force density $f_\beta $ in the Yang-Mills flux tube;
\begin{align}
    \frac{\partial}{\partial x^\alpha} T_{\alpha \beta} = f_\beta &= F_{\beta \lambda }  \frac{\partial}{\partial x^\alpha}F_{\alpha \lambda}\nonumber\\
     &- \frac{F_{\mu \lambda}}{2} (\partial_\beta F_{\mu \lambda} + \partial_\mu F_{\lambda \beta} + \partial_\lambda F_{\beta \mu}).
    \label{Fs}
\end{align}

Identifying the magnetic currents,
\begin{equation}
     J_\alpha^{\rm mag} \equiv \frac{1}{2} \epsilon_{\alpha \beta\mu \lambda} \frac{\partial F_{\mu \lambda}}{\partial x^\beta},~~~~~~~( \epsilon_{4123} = 1) \;,
     \label{Jmag2}
 \end{equation}
 we can express the force density \eqref{Fs}
 as the sum of an electric Lorentz force density and a magnetic Lorentz force density
\begin{align}
     f_\beta  = -F_{\beta \lambda} J^{\rm el}_\lambda  - F_{\mu \lambda} ~\frac{1}{2} \epsilon_{\alpha \beta\mu \lambda} J^{\rm mag}_\alpha, 
    \label{fdensity}
\end{align}
  corresponding to a flux tube comprised of gauge invariant electric and magnetic currents measured in our simulations. This description emerges from using the divergence of the Maxwell stress tensor
to calculate forces, without requiring that the field tensor $F_{\mu \nu}$ satisfy Maxwell's equations.

To test this picture we perform new simulations of the electric field $\vec{E}$, and numerically evaluate the magnetic current density $\vec{J}_{\rm mag}$ and the electric charge density $\rho_{\rm el}$ in the flux tube.

If $F_{\alpha \beta}$ had been the field tensor of electrodynamics, the second term on the RHS of \Cref{Fs} would vanish because of the homogeneous Maxwell equations, and the first term would be fixed by the inhomogeneous Maxwell equations in terms of the electric current density of the charged matter. The right hand side of \Cref{Fs} then would reduce to the expression for the Lorentz force density.

For the Yang-Mills flux tube, $F_{\alpha \beta}$ (see \Cref{connected2}) is the Maxwell-like field tensor measured in our simulations and the derivatives on the right hand side of \Cref{Fs} define  the  electric current density $J_\beta^{\rm el}$ and the magnetic  current density $J_\beta^{\rm mag}$ generated by the field $F_{\alpha \beta}$:
\begin{itemize}
    \item[$\bullet$] $J^{\rm el}_4$ and $J^{\rm mag}_4$ are the electric charge density $\rho_{\rm el}$ and magnetic charge density $\rho_{\rm mag}$, respectively;
    \item[$\bullet$] $J^{\rm el}_i$ and $J^{\rm mag}_i$, $i = 1,2,3$ are the  components of the vector electric current density $\vec{J}_{\rm el}$ and the vector magnetic current density $\vec{J}_{\rm mag}$, respectively.
\end{itemize}

The current densities expressed in terms of the electric components $E_k = F_{4k}$ and the magnetic components $B_i = \frac{1}{2} \epsilon_{ijk} F_{jk}$ of the field tensor $F_{\mu \nu}$ have the form
 \begin{equation}
    \rho_{\rm el}  = \vec{\nabla} \cdot \vec{E}\;, \;\;\;\;\; \vec{J}_{\rm el} = \vec{\nabla} \times \vec{B} - \frac{\partial \vec{E}}{\partial x_4}\;,
    \label{fieldderivs2}
\end{equation}
\begin{equation}
    \rho_{\rm mag} = \vec{\nabla} \cdot \vec{B}\;, \;\;\;\;\;\vec{J}_{\rm mag} = \vec{\nabla} \times \vec{E} - \frac{\partial \vec{B}}{\partial x_4}\;,
    \label{fieldderivs1}
 \end{equation}
while \Cref{fdensity} for the
Lorentz force density $\vec{f}$ assumes the  form 
\begin{equation}
    \vec{f} = \rho_{\rm el} ~\vec{E} + \vec{J}_{\rm mag}  \times \vec{E} - \rho_{\rm mag}  \vec{B} - \vec{J}_{\rm el} \times \vec{B}\;.
\label{vecfdensity}
\end{equation}
Since the magnetic field $\vec{B}$ simulated in the static flux tube vanishes within statistical error, \Cref{vecfdensity}
simplifies to
\begin{equation}
    \vec{f} = \rho_{\rm el}~ \vec{E} + \vec{J}_{\rm mag} \times \vec{E}.
   \label{vecf1}
\end{equation}
Furthermore, since $\vec{\nabla} \times \vec{E}^C =0$, the perturbative Coulomb-like field $~\vec{E}^{\rm C}$ does not contribute to the magnetic current density. Therefore,
\begin{equation}
\vec{J}_{\rm mag} = \vec{\nabla} \times\vec{E}^{NP}\;.
\label{Jmagandrhoel}
\end{equation}

We will see from our simulations that the electric charge density $\rho_{\rm el} = \vec{\nabla} \cdot \vec{E}$ is significantly different from zero only at the positions of the quark sources.

Consequently, the force density $\vec{f}$ interior to the flux tube $(0 \le x_l \le d)$ is the magnetic Lorentz force density 
\begin{equation}
 \vec{f} = \vec{J}_{\rm mag} \times \vec{E}^{NP} = - \hat{e}_y ~J^{\rm mag}_z (x_l, x_t)\ E^{NP}_x (x_l, x_t)
  \label{vecf}
  \end{equation}
directed towards the flux tube axis.

We can imagine cutting the flux tube along any plane containing its axis. We then calculate  the force $\vec{F}$ on one half of the flux tube cut by the plane.

Making use of rotational symmetry, we introduce polar coordinates $(x_t, \theta)$ in the transverse $(y,~z)$ plane.  $\theta$ is the angle between the $y$-axis and $\hat{e}_{x_t}$, the unit vector in the direction of $\vec{f}$ ; {\it{i. e.} }, $y = x_t ~\mathrm {cos} \theta,~~~z=x_t~ \mathrm{sin} \theta$.

The force $\vec{F}$ on this half flux tube
is then given by:
\begin{equation}
    \vec{F} = -\int\limits_0^d \mathrm{d}x_l~\int\limits_0^\infty \mathrm{d} x_t~x_t \int\limits_{-\pi/2}^{\pi/2} \mathrm{d}\theta  (\cos \theta ~\hat{e}_y + \sin \theta~ \hat{e}_z) f (x_l, x_t),
\end{equation}
where $f(x_l, x_t)$ is the magnitude of the Lorentz force density, \Cref{vecf}.
The integral $\int \cos \theta \mathrm{d} \theta =2$, and the integral  $\int \sin \theta \mathrm{d} \theta 
~=0$, so that the expression for the force $\vec{F}$ yields the result
\begin {equation}
    \vec{F} =~- \hat{e}_y \left( 2 \int_0^d \mathrm{d}x_l \int_0^\infty \mathrm{d}x_t\, x_t f(x_l, x_t)\right)
    \label{vecF}
\end{equation}
directed toward the flux tube axis.

Integrating the force density $\vec{f}$ over one half of the cut flux tube yields the force $\vec{F}$ on that half; integrating $\vec{f}$ over the other half of the tube yields an equal and opposite force on that half, pushing the two halves together. The flux tube is thus confined in the transverse direction by this `squeezing' force $\vec{F}$. 

We use our numerical simulations of the connected correlator to evaluate the integrand in~\Cref{vecF} determining the magnitude of the magnetic Lorentz force density  in the half flux tube.  Figure~\ref{fig:integrand} shows the magnitude of this force density generated by magnetic currents for three values of $\beta$.

In Section (\ref{sec:Force}) we show the confining force $\vec{F}$ obtained from the integration \Cref{vecF} over the half flux tube is compatible within systematic errors to the value of the string tension $\sigma$ calculated from the integral \Cref{stress} of $T_{xx}$ over the midplane between the quarks, checking the consistency of these two representations of confinement.

\section {Direct simulation of the magnetic current density}
\label{sec:magnetic}
The possible presence of magnetic currents in SU(3) lattice gauge theory theory was pointed out in~\Onlinecite{Skala:1996ar}, where it was noted that, in contrast to the magnetic monopoles in U(1) lattice gauge theory, the magnetic currents in non-Abelian lattice gauge theory need not be quantized.

In this paper we evaluate the magnetic current density  numerically, using our lattice measurements of the connected correlator.

To evaluate
\begin {equation}
J^{\rm mag}_z = \frac{\partial E_x}{\partial y} - \frac{\partial E_y}{\partial x} \;,
\label{Jz}
\end{equation}
numerically, we replace the field derivatives in \Cref{Jz} by the corresponding differences of expectation values of fields measured in our lattice simulations of the connected correlator; {\em i.e.} (see \Cref{sec:correlator}, \Cref{fig:op_W} and \Cref{connected1,connected2}), 
\begin {align}
&\frac{\partial E_x}{\partial y} \rightarrow \langle E_x (x_l, x_t +1)\rangle_{q \bar{q}} -  \langle E_x (x_l, x_t)\rangle_{q \bar{q}} ;&
\label{discretederivs1}
\\
&\frac{\partial E_y}{\partial x} \rightarrow \langle E_y (x_l +1, x_t)\rangle_{q \bar{q}} -  \langle E_y (x_l, x_t)\rangle_{q \bar{q}}.&
\label{discretederivs2}
\end{align}

Replacing \Cref{discretederivs1} and \Cref{discretederivs2} in \Cref{Jz} yields the magnetic current distributions $J^{\rm mag}_z$ corresponding to each of the $q \bar{q}$ separations for which we carried out simulations of the color field distributions. 

\begin{figure}[htb]
\centering
\includegraphics[width=0.7\linewidth,clip]{{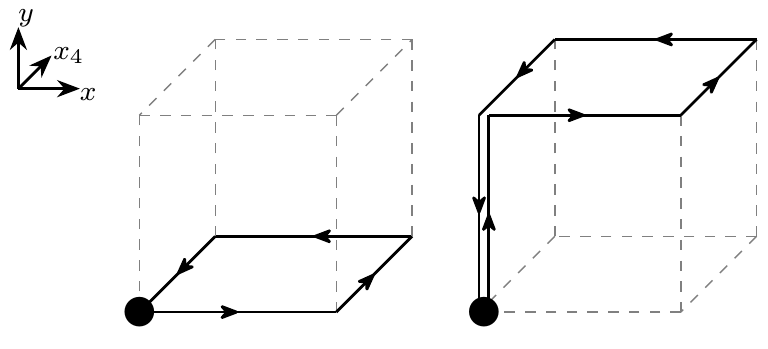}}
\caption{A pair of paths of closed loops defining a contribution to  the magnetic current in the $z$ direction, according to~\cite{Skala:1996ar}.}
\label{fig:cubes}
\end{figure}

Before presenting the numerical results of these measurements, we now show that they can be identified as measurements of expectation values of loops constructed from plaquettes lying on opposite faces of an $(x, y, x_4)$ cube, as depicted in \Cref{fig:cubes}, and consequently are measurements of the flux of  magnetic current out of the cube, as proposed in~\Onlinecite{Skala:1996ar}.

The black dot in \Cref{fig:cubes} stands for the Wilson loop $W$ and connecting Schwinger lines $L$ and $L^+$ attached to the plaquette $U_{41} (x_l, x_t)$ in \Cref{fig:op_W}. The loop on the left in \Cref{fig:cubes} is then a condensed representation of \Cref{fig:op_W} measuring the longitudinal electric field  $\langle E_x (x_l, x_t)\rangle_{q \bar{q}}$.
In the loop on the right in \Cref{fig:cubes}, the plaquette  $U_{41} (x_l, x_t +1)$, translated by one unit in the $y$ direction and measuring $\langle E_x (x_l, x_t+1)\rangle_{q \bar{q}}$, is attached to Schwinger lines extending  $L$ and $L^+$ by single links $U_2 (x_l, x_t)$ and $U_2^+ (x_l, x_t)$ in the $y$ direction.
 
\Cref{discretederivs1} for $\frac{\partial E_x}{\partial x}$ is then the expectation value of the difference of two loops constructed from plaquettes $U_{41} $ lying on opposite $xx_4$ faces of the $(x, y, x_4)$ cube in \Cref{fig:cubes}. 
 
Likewise, \Cref{discretederivs2} expresses $\frac{\partial E_y}{\partial x}$ as the expectation value of the difference of a corresponding pair of loops constructed from plaquettes $U_{42}$ lying on opposite $yx_4$ faces of the cube, so that $J^{\rm mag}_z$ can be expressed as the sum of contributions from the $yx_4$ and $xx_4$ faces of the  cube.
 
The  term $-\frac{\partial B_z}{\partial x_4}$  in \Cref{fieldderivs1} gives the contribution to $J^{\rm mag}_z$  from plaquettes $U_{12}$ on the $xy$ faces of the `magnetic current' $(x, y, x_4)$ cube.
 
We can generalize \Cref{Jz}, \Cref{discretederivs1} and \Cref{discretederivs2}, using the 4-dimensional form \Cref{Jmag2} for the magnetic current density $J_\alpha^{\rm mag}$.

Making the replacements
\begin{equation}
    \frac{\partial F_{\mu \lambda}}{\partial x^\beta}~~ \rightarrow ~~\frac{1}{a} \left(\langle F_{\mu \lambda} (x + a \hat{\beta})\rangle_{q \bar{q}} - \langle F_{\mu \lambda} (x)\rangle_{q \bar{q}} \right )
    \label{Jgeneralized}
\end{equation}
in \Cref{Jmag2} yields the magnetic current density $\langle J_\alpha^{\rm  mag}\rangle_{q \bar{q}}$ measuring the flux of magnetic current out of a $(\hat{\beta}, \hat{\mu}, \hat{\lambda})$ magnetic current cube (the $(y, x_4, x)$ cube in \Cref{fig:cubes}.
    
The translated plaquette $U_{\lambda \mu} (x + a \hat{\beta})$ is attached to single links $U_a (\hat{x}) \equiv e^{i a A_\beta}$ and $U_\beta^+ (x) \equiv e^{-i a A_\beta}$, accounting for the extensions of the attached Schwinger lines, so that:
\begin {align}
&g a^2 (F_{\mu \lambda} (x + a \hat{\beta}) - F_{\mu \lambda} (x)) \nonumber \\&= e^{ i a A_\beta} U_{ \mu \lambda} (x + a \hat{\beta}) e^{-i a A_\beta} - U_{ \mu \lambda} (x)\;.
    \label{extensions}
\end{align}
Inserting \Cref{Jgeneralized} in \Cref{Jmag2} yields
    \begin {equation}
    \langle J_\alpha^{\rm mag} (x) \rangle_{q \bar{q}}~ =~ \frac{1}{2} \epsilon_{\alpha \beta    \mu \lambda} \frac{\langle F_{  \mu \lambda} (x + a \hat{\beta}) - F_{\mu \lambda  } (x) \rangle_{q \bar{q}}}{a}\;.
    \label{jmagexpect}
\end{equation}
The sum over $\hat{\beta}, \hat{\mu}, \hat{\lambda} $ is a sum over the six faces of the $\hat{\beta}, \hat{\mu}, \hat{\lambda}$ magnetic current cube contributing to the $\alpha$ component of the magnetic current density.
    
If we were to first take the continuum limit $a \rightarrow 0$ for fixed $q \bar{q}$ separation in lattice units (na\"ive continuum limit), \Cref{extensions,jmagexpect} would reduce to the Bianchi identity:
    \begin {equation}
    \lim_{a \rightarrow 0} \langle J^{\rm mag}_\alpha (x)\rangle_{q \bar{q}}~~ \rightarrow ~~\langle \frac{1}{2} \epsilon_{\alpha \beta \mu \lambda} (D_\beta F_{ \mu \lambda} )\rangle_{q \bar{q}} ~=~0\;,
    \label{Bianchi}
\end{equation}
where
\begin {equation}
    D_\beta F_{ \mu \lambda} \equiv \frac {\partial F_{ \mu \lambda}}{\partial x^\beta} + i [A_\beta, F_{ \mu \lambda}]\;. 
    \label{Dbeta}
\end{equation}

Instead, we simulated $\langle J_\alpha^{\rm mag} (x) \rangle_{q \bar{q}}$, reducing the lattice spacing, keeping the quark-antiquark separation $d$ fixed in physical units.
We  have carried out numerical simulations for pure SU(3) gauge theory on a $48^4$ lattice for three values of the gauge coupling $\beta$, all of which correspond to the quark-antiquark separation $d = 0.511$~fm  (See \Cref{betavalues}).
%
\section{Lattice setup and smearing procedure}
\label{sec:setup}

\begin{figure}[htb!]
\centering
\subfigure[$E_x(x_l = d/2,  x_t = d/2)$]%
{\label{fig:smearing_Ex}\includegraphics[width=\columnwidth,clip]{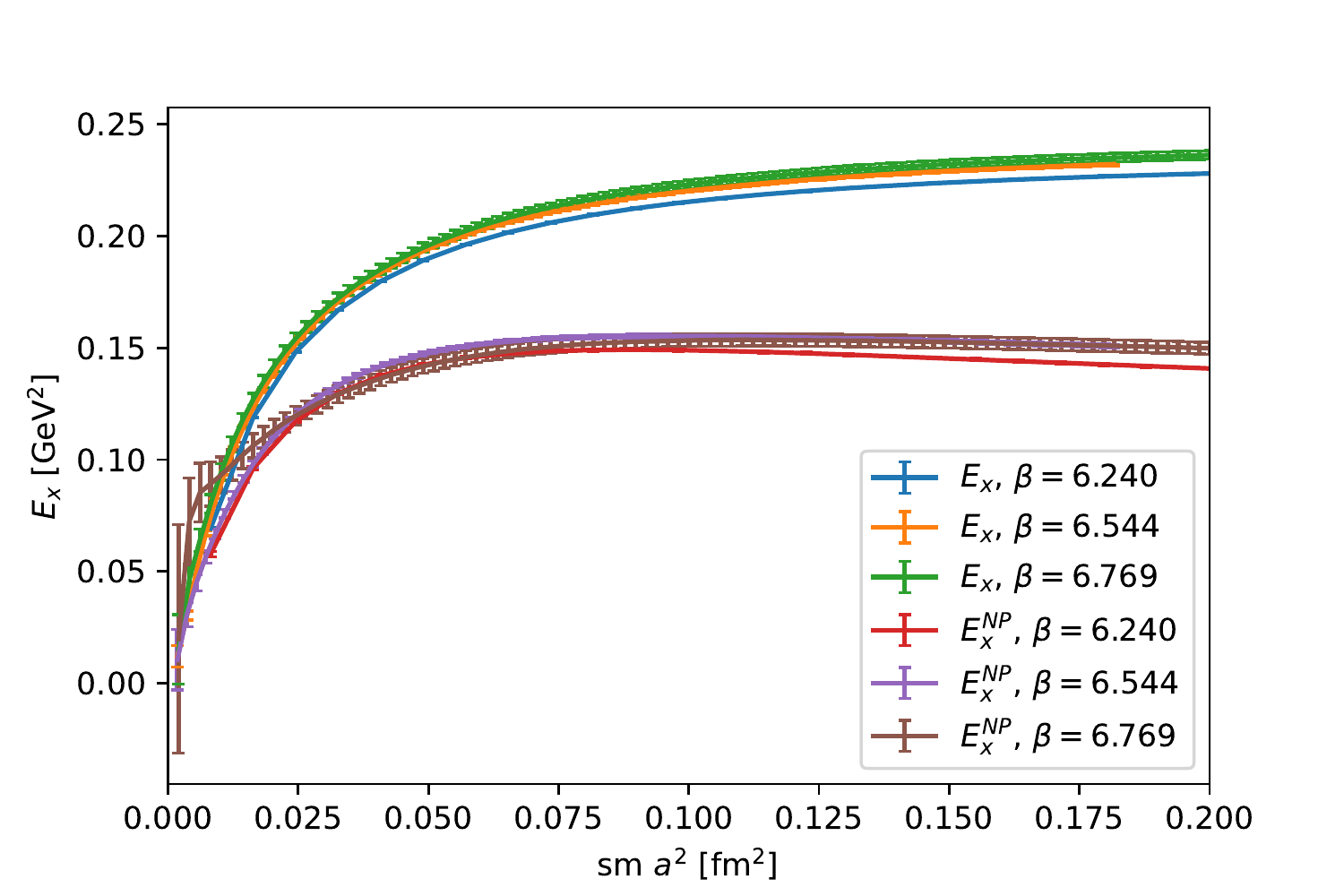}}
\subfigure[$\rho_{\rm el} (x_l = d / 4, x_t = d / 4)$] 
{\label{fig:smearing_divE}\includegraphics[width=\columnwidth,clip]{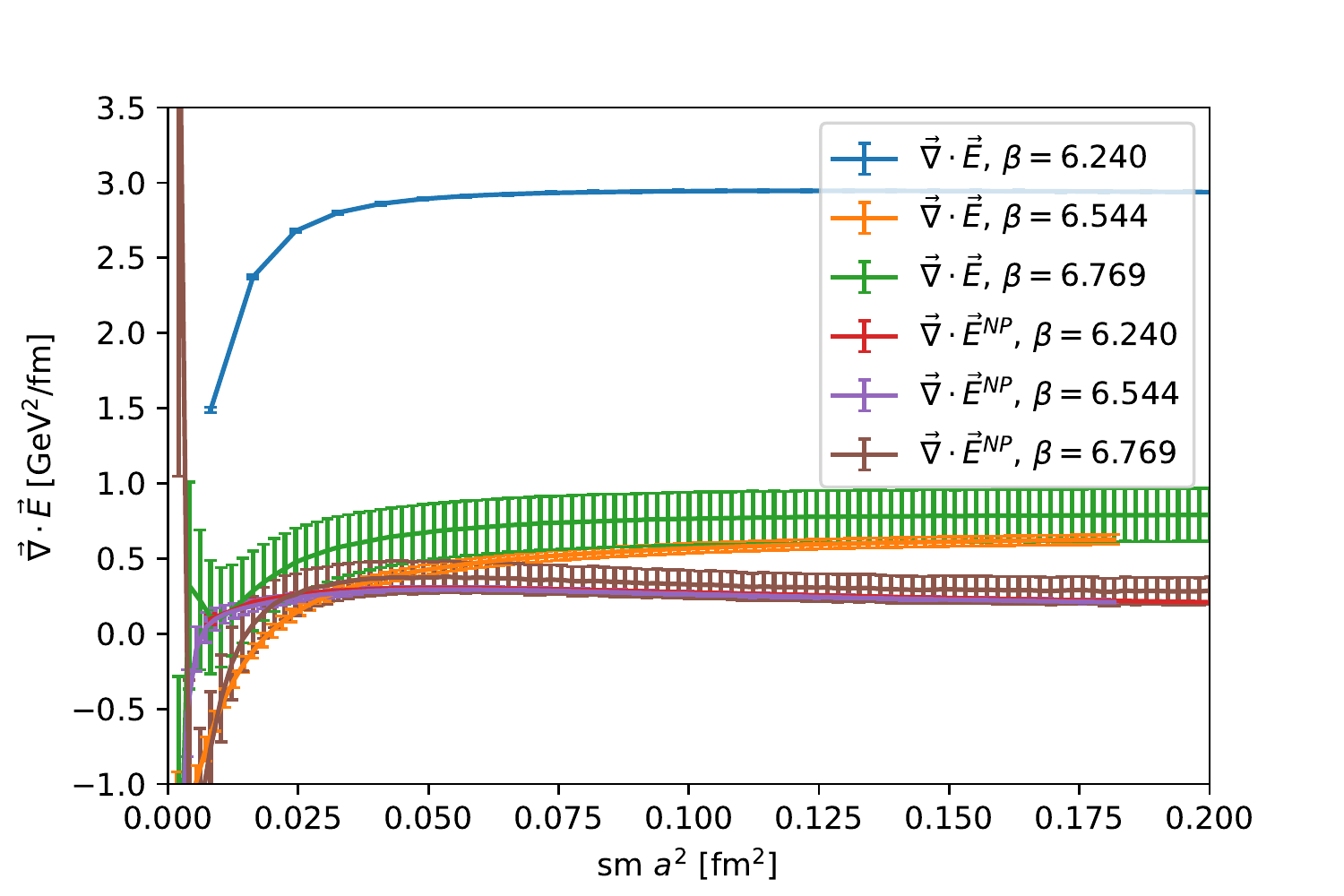}}
\subfigure[$(\vec{J}_{\rm mag})_z (x_l = d / 4, x_t = d / 2)$] 
{\label{fig:smearing_curlE}\includegraphics[width=\columnwidth,clip]{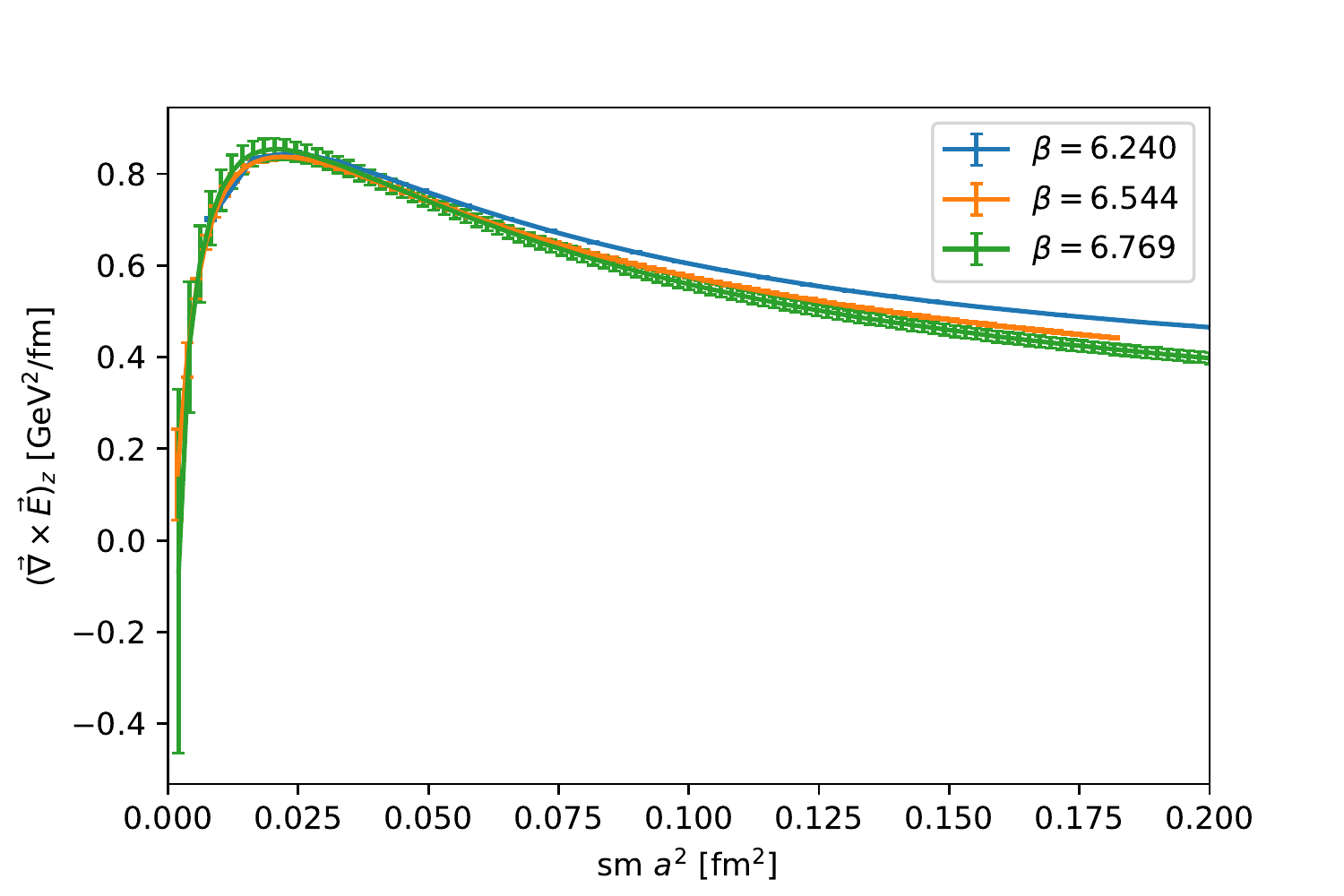}}
\caption{Full and non-perturbative longitudinal electric field  
at $x_l = x_t = d / 2$~(\ref{fig:smearing_Ex}),
full and non-perturbative electric charge density
at $x_l = x_t = d / 4$~(\ref{fig:smearing_divE}), and
magnetic current density
at $x_l = d / 4$, $x_t = d / 2$~(\ref{fig:smearing_curlE}) as functions of smearing.}
\label{fig:smearing}
\end{figure}

\begin{table*}[tb]
\begin{center} 
  \caption{Summary of the numerical simulations.}
  \label{betavalues}
\newcolumntype{Y}{>{\centering\arraybackslash}X}
\begin{tabularx}{0.9\linewidth}{@{}YYYYYYY@{}}
\toprule
lattice & $\beta=6/g^2$ & $a(\beta)$ [fm]  & $d$ [lattice units]   & $d$ [fm] & statistics\\ \midrule
$48^4$ & 6.240 & 0.0639 & 8 & 0.511 & 268 \\
$48^4$ & 6.544  & 0.0426 & 12 & 0.511 & 508 \\
$48^4$ & 6.769 & 0.0320  & 16 & 0.511 & 303 \\
\bottomrule 
\end{tabularx} 
\end{center}
\end{table*}
We measured the color fields, as defined in \Cref{connected1},  generated by a quark-antiquark pair separated by distance $d$. For each value of the gauge coupling $\beta$ in \Cref{betavalues}  measurements were performed every 25 upgrades of the gauge configuration. The  value of $\beta$ was determined for each value of $d$[lattice units] such that  $d\approx0.511 {\text{ fm}}$ in physical units, where the physical scale for the lattice spacing $a(\beta)$ was set according to \Onlinecite{Necco:2001xg}:
\begin{align}
\label{NSscale}
& a(\beta) = r_0 \! \times \! \exp\left[c_0+c_1(\beta\!-\!6)+c_2(\beta\!-\!6)^2+c_3(\beta\!-\!6)^3\right], \nonumber\\
& r_0 = 0.5 \; {\rm fm}, \nonumber\\
& c_0=-1.6804 \,,  c_1=-1.7331 \,,\nonumber\\
& c_2=0.7849 \,, c_3=-0.4428 \,,
\end{align}
for all $\beta$ values in the range $5.7 \le \beta \le 6.92$.
In this scheme (see Eq. (3.5) in \Onlinecite{Necco:2001xg}) we have, for the value of the square root of the string tension, \mbox{$\sqrt{\sigma} \approx 0.464 \, {\textrm{GeV}}$}.

The connected correlator defined in \Cref{connected1} exhibits large
fluctuations at the scale of the lattice spacing, which are responsible
for small signal-to-noise ratio. To extract the physical information carried
by fluctuations at the physical scale (and, therefore, at large distances
in lattice units) we smoothed out configurations by a {\em smearing}
procedure.

Our setup consisted of one step of 4-dimensional hypercubic smearing~\cite{Hasenfratz:2001hp} on the temporal links (HYPt), with smearing parameters $(\alpha_1,\alpha_2,\alpha_3) = (1.0, 1.0, 0.5)$, and $N_{\rm HYP3d}$ steps of hypercubic smearing  restricted to the three spatial directions (HYP3d) with $(\alpha_1^{\text{HYP3d}},\alpha_3^{\text{HYP3d}}) = (0.75, 0.3)$.

The operator in \Cref{connected1} which defines the color field strength tensor undergoes a non-trivial renormalization~\cite{Battelli:2019lkz}, which depends on $x_t$. As discussed in \Onlinecite{Baker:2018mhw,Baker:2019gsi}, comparing our results with those in \Onlinecite{Battelli:2019lkz} we argued that smearing behaves as an effective renormalization.

\Cref{fig:smearing_Ex} shows that, at  $x_t = x_l = d/2$, both the full and the non-perturbative longitudinal electric fields display plateaus as a function of the amount of smearing - quantified as the number of smearing steps times the squared lattice spacing - at all considered values of $\beta$ (and the same holds true at all values of $x_t$). The value of the longitudinal component $E_x$ for the full field shows no sign of degradation of the signal within the explored range of smearing steps, while for the longitudinal component $E_x^{NP}$ there is a very wide maximum, followed by a decrease of the smeared value when the number of smearing steps is further increased.
We extract $x_l$- and $x_t$-dependent optimal amounts of smearing for the field based on the position of the above mentioned maximum.

\begin{figure}[htb]
\centering
\includegraphics[width=\columnwidth,clip]{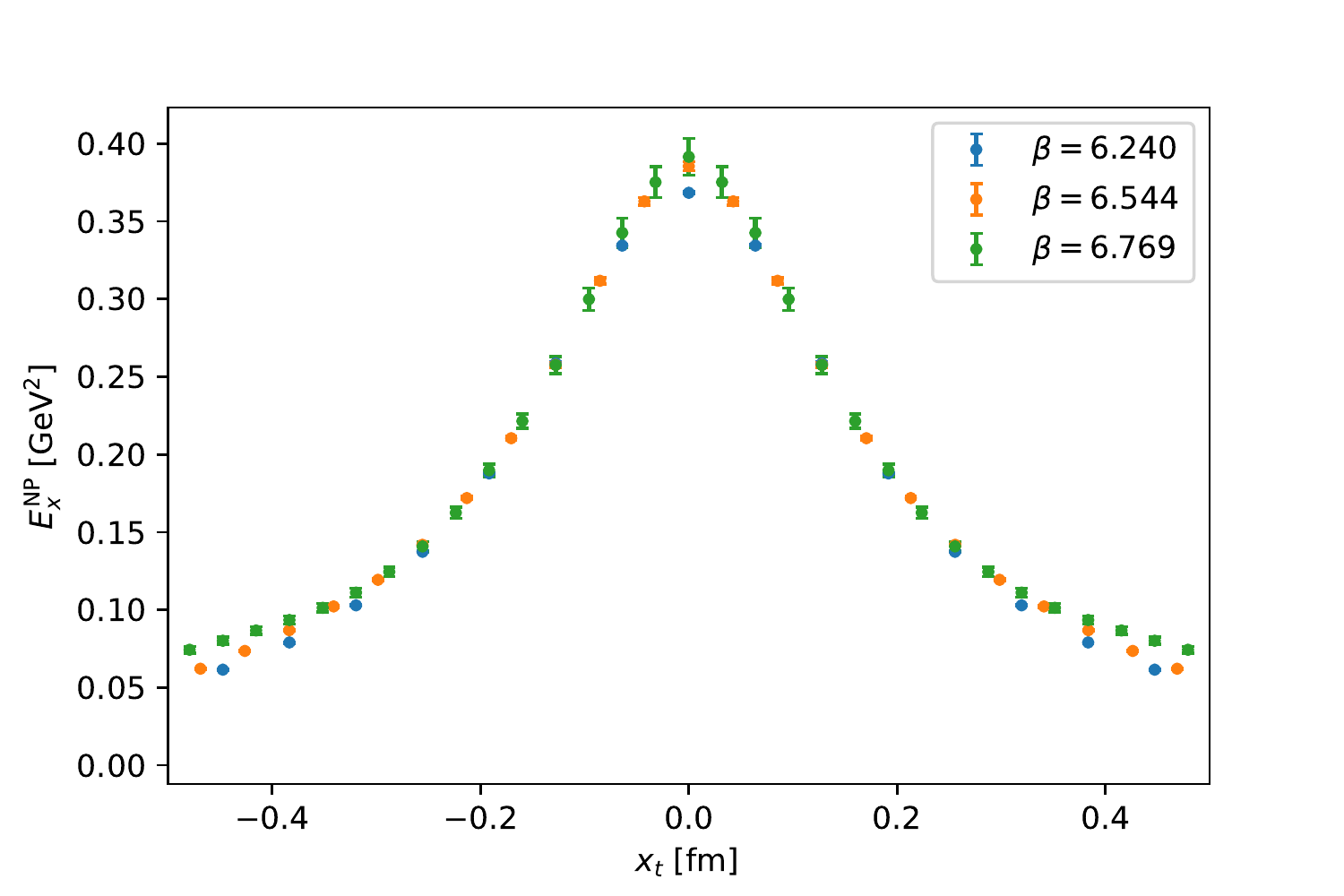}
\caption{Continuum scaling of the longitudinal component of the non-perturbative field $E_x^{\rm NP}$ 
at a transverse plane $x_l = d/4 \approx 0.128$~fm.}
\label{fig:scalingField}
\end{figure}

The smearing procedure can also be validated {\em a posteriori} by the observation of continuum scaling, that is by checking that fields obtained in the same {\em physical} setup, but at different values of the coupling and of the optimal number of smearing steps, are in good agreement in the range of gauge couplings used.
This is seen, for the non-perturbative field, in~\Cref{fig:scalingField}, where
$E^{NP}_x (x_t)$   at a transverse plane $x_l = d/4$ is plotted for three values of $\beta$. 
 
When it comes to the behavior {\em versus} the number of spatial smearing steps of the magnetic current density (see \Cref{fig:smearing_curlE}) the situation is somewhat different: We observe a significant degradation of the signal after some small amount of smearing.
Smearing still plays a role in improving the signal-to-noise ratio and eliminating short-distance UV fluctuations, but a small amount of smearing seems sufficient.

\begin{figure}[t]
	\centering
	\label{fig:cs2}\includegraphics[width=0.9\linewidth]{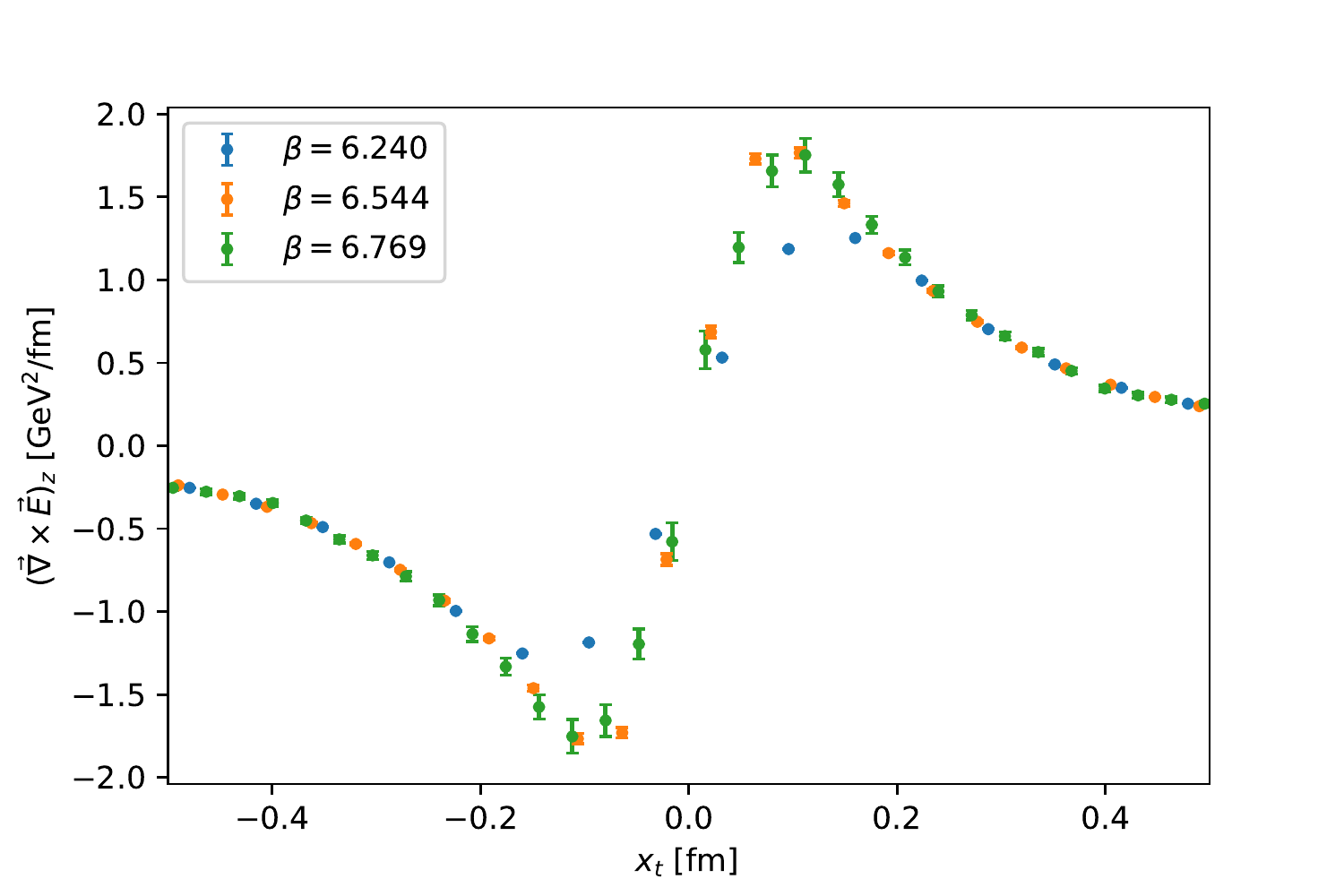}
	\hfill
	\caption{\label{fig:contScalingCurl}Continuum scaling for the
    magnetic current $(\vec{\nabla} \times \vec{E})_z$ extracted at the (position dependent) optimal value of smearing at a transverse plane $x_l = d/4 \approx 0.128$~fm.}
\end{figure}

Also in this case, our choice of identifying the optimal amount of smearing by the maximum in $(\vec{\nabla} \times \vec{E})_z$ as a function of the amount of smearing is validated by how nicely results obtained at different values of $\beta$ scale at the maximum, as opposed to at larger amounts of smearing (see  \Cref{fig:contScalingCurl}).

\begin{figure}[htb]
   \centering
   \includegraphics[width=1.\linewidth,clip]{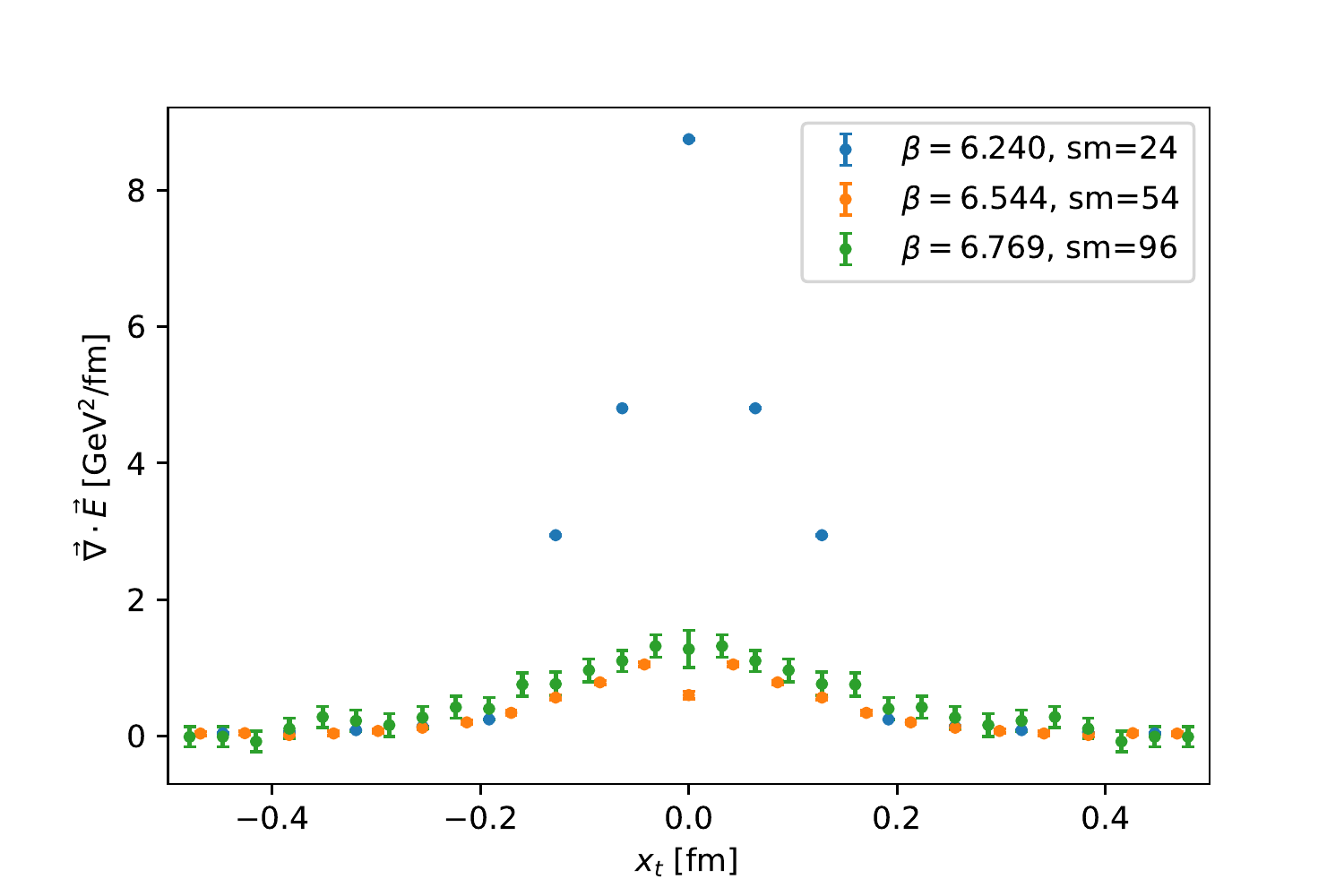}
   \caption{Scaling of the divergence of full field at a transverse plane $x_l = d/4 \approx 0.128$~fm}
   \label{fig:divE_scaling}
\end{figure}

Finally, the behavior {\em versus} smearing of the full and non-perturbative electric charge density (see \Cref{fig:smearing_divE}) is indicative of the fact that these quantities will vanish when going to the continuum limit (see also \Cref{fig:divE_scaling}).
In this case we fix the amount of smearing by the plateau displayed by $\vec{\nabla} \cdot \vec{E}$ at our smallest $\beta$ and then we scale that amount of smearing on our finer lattices by taking into account the diffusive nature of the smearing process.

\section {Numerical results}\label{sec:numerical}

\subsection {Numerical Evaluation of $\rho_{\rm el}$}
\label{sec:num_rho}

As shown in section (\ref{sec:setup}), the behavior under smearing of the divergence of both full and non-perturbative fields does not show a maximum and a further decay, but rather approaches a plateau, which in most cases is close to zero. 

Additionally, the scaling of the smeared divergences is worse than for other observables, suggesting that the values we obtain have large contributions from discretization errors. 
In this case we have to use another approach to the choice of the optimal smearing step, namely fixing it at a value that lies in the plateau region for all lattice points.

\begin{figure}[htb]
   \centering
   \includegraphics[width=1.\linewidth,clip]{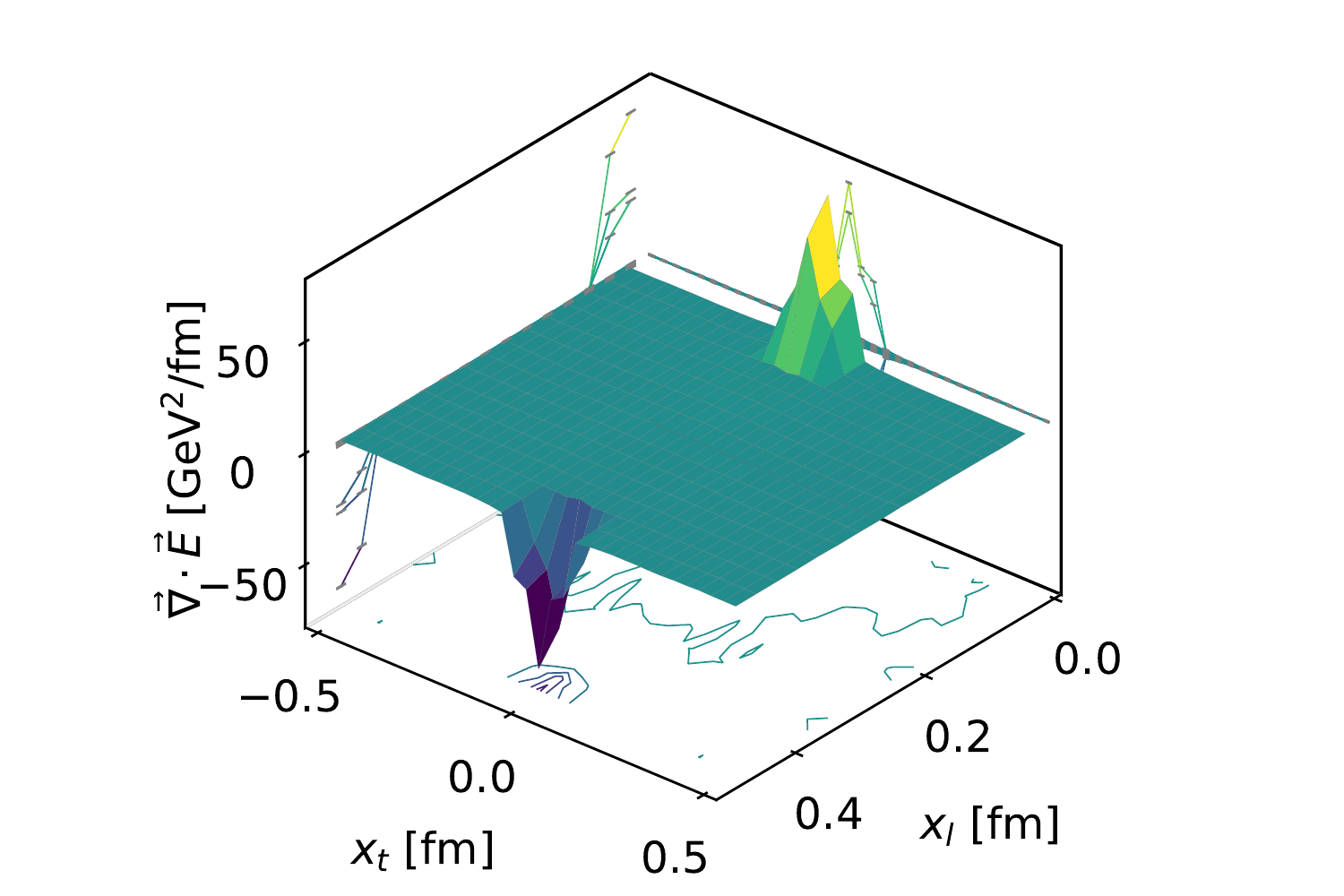}
   \caption{Divergence of the full field produced by a quark-antiquark pair separated by $d=0.51$~fm, corresponding to $\beta = 6.769$ after 96 smearing steps.}
   \label{fig:divE}
\end{figure}

\Cref{fig:divE} shows $\vec{\nabla} \cdot \vec{E}$, the divergence of the full field $\vec{E}$ 
for $d=0.51$~fm at $\beta=6.769$, after 96 smearing steps.
As one can clearly see, the values of the divergence are significantly different from zero only at distances of about two lattice steps away from the sources. 

The visible peaks of the divergence of the full field are removed by the subtraction of its perturbative part, so the non-perturbative charge density becomes negligible. 

\subsection {Numerical Evaluation of $\vec{J}_{\rm mag}$}
\label{sec:num_jmag}

\begin {figure}[htb]
\centering
   \subfigure[$\beta = 6.240$]%
             {\label{fig:jBeta240}\includegraphics[width=1.\linewidth,clip]{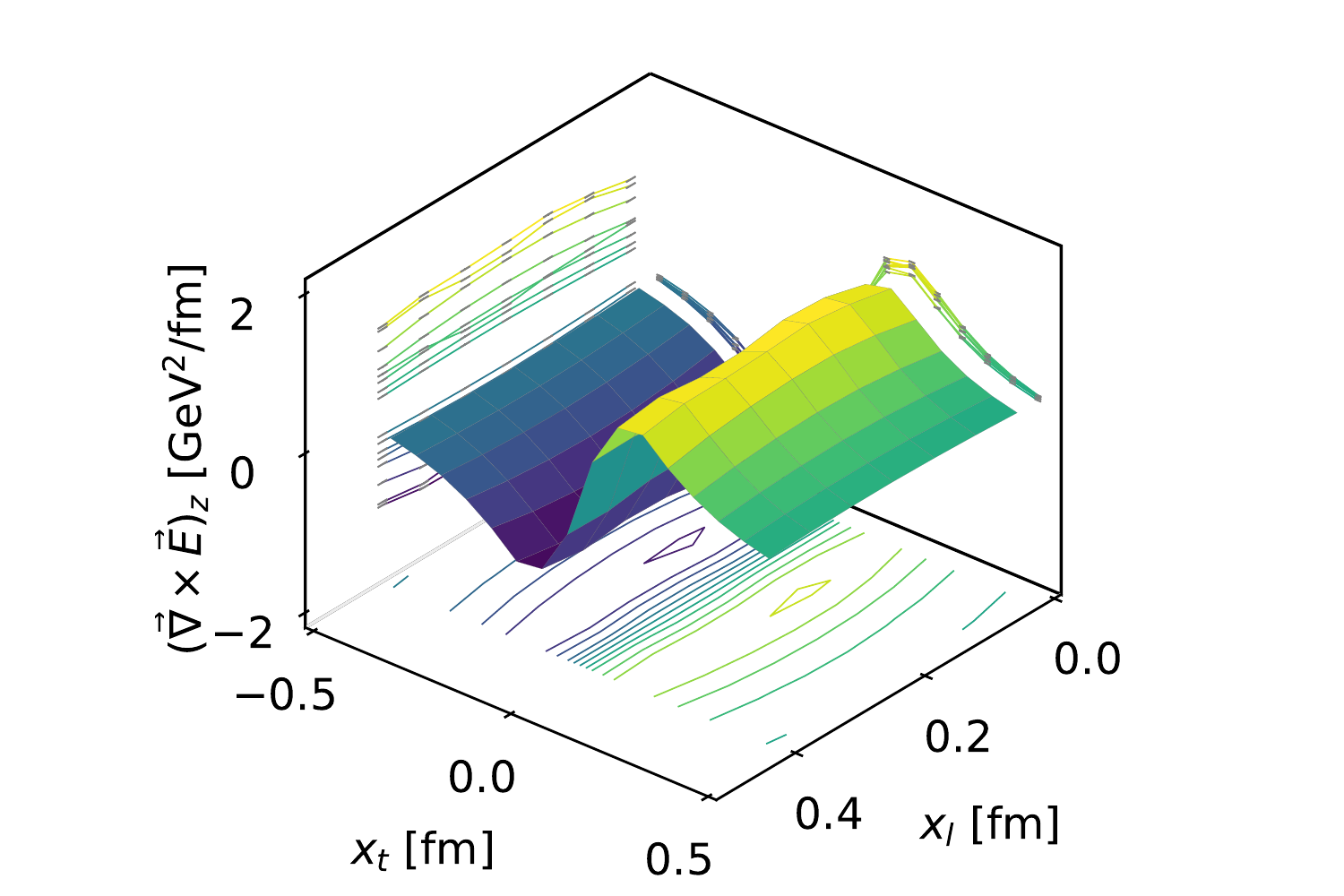}}
   \subfigure[$\beta = 6.769$]%
             {\label{fig:jBeta769}\includegraphics[width=1.\linewidth,clip]{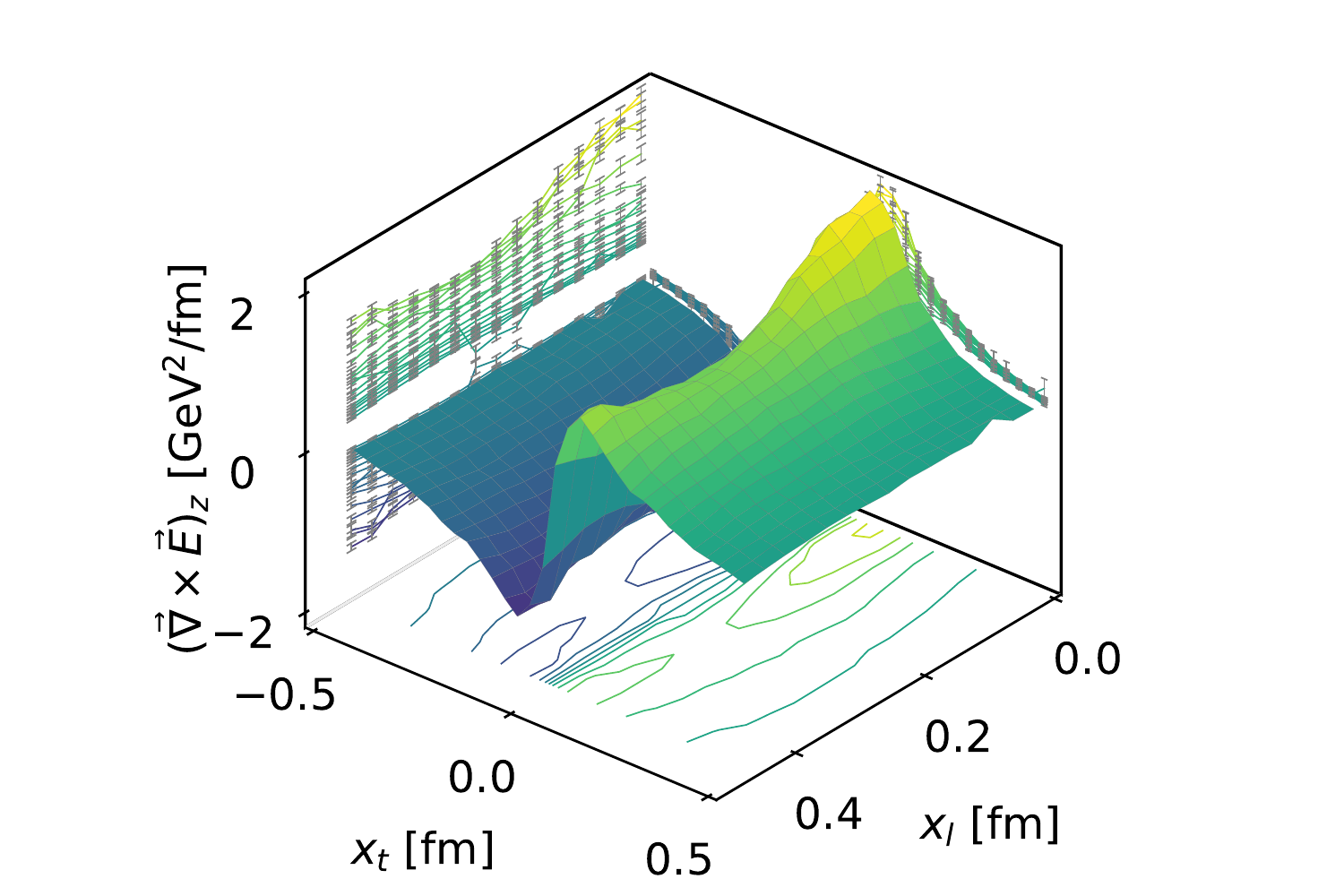}}
\caption{Magnetic current $(\vec{\nabla} \times \vec{E})_z$
created by a quark-antiquark pair separated by $d=0.51$~fm, for $\beta = 6.240$~(\ref{fig:jBeta240}) and $\beta = 6.769$~(\ref{fig:jBeta769}). }
\label{fig:rotE}
\end{figure}

As seen in~\Cref{fig:rotE} the magnetic current density $\vec{J}_{\rm mag} (x)$, in contrast to the  electric charge density  $\rho_{\rm el} (x)$, is manifest throughout the whole length of the flux tube.
$\vec{J}_{\rm mag} (x)$ points in the direction $\hat{e}_{\theta}$, circulating about the flux tube axis.

We use the procedure defined in the previous section, to extract the amount of smearing resulting in the local maximum of $\vec{J}_{\rm mag} (x)$ at each point $(x_l, x_t)$.
\Cref{fig:rotE} shows the dependence of the magnetic current density on $x_l$ and $x_t$ for $d = 0.51$~fm and two values of $\beta$.

The continuum scaling behavior of the magnetic current density is shown in~\Cref{fig:contScalingCurl}.
One can see that all three scales give compatible values of the current density except for 
the smallest $\beta$ at the points close to $x_t = 0$.

Note that,  since the perturbative field is defined explicitly as that for which the curl is equal to zero, the non-perturbative  current density, as opposed to the charge density, is exactly equal to the full current density, and thus is not dependent on the conjecture that the non-perturbative field is purely longitudinal.

\subsection {Calculation of the confining force $\vec{F}$ and the string tension $\sigma$}
\label{sec:Force}
\begin{table*}[htb]
\begin{center} 
\caption{Summary of the numerical determination of the (square root of the) force, \Cref{vecF} and the (square root of the) stress tensor \Cref{eq:stringtension}; the last
four columns give the edges of the integration domains. See the text for how the assessment of systematics on $\sqrt{F}$ is performed.}
\label{force}
\newcolumntype{Y}{>{\centering\arraybackslash}X}
\begin{tabularx}{0.9\linewidth}{@{}YYYYYYY@{}}
\toprule
$\beta$ & $\sqrt{F}$ [GeV] & $\sqrt{\sigma}$ [GeV] & $x_{l,{\rm min}}$ [fm]&  $x_{l,{\rm max}}$ [fm]& $x_{t,{\rm min}}$ [fm]&  $x_{t,{\rm max}}$ [fm]\\ \midrule
6.240 & 0.4859(4)${}^{+645}$         & 0.4742(12) & 0.031951 & 0.415364 & 0. & 1.022434 \\
&&&&&&\\
6.544 & 0.5165(8)${}^{+611}_{-214}$  & 0.4692(16) & 0.021312 & 0.447542 & 0. & 0.511477 \\
&&&&&&\\
6.769 & 0.5297(22)${}^{+547}_{-322}$ & 0.4672(49) & 0.015976 & 0.46331  & 0. & 0.511238 \\
\bottomrule
\end{tabularx} 
\end{center}
\end{table*}
\begin {figure*}[htb]
\centering
\includegraphics[width=0.75\columnwidth,clip]{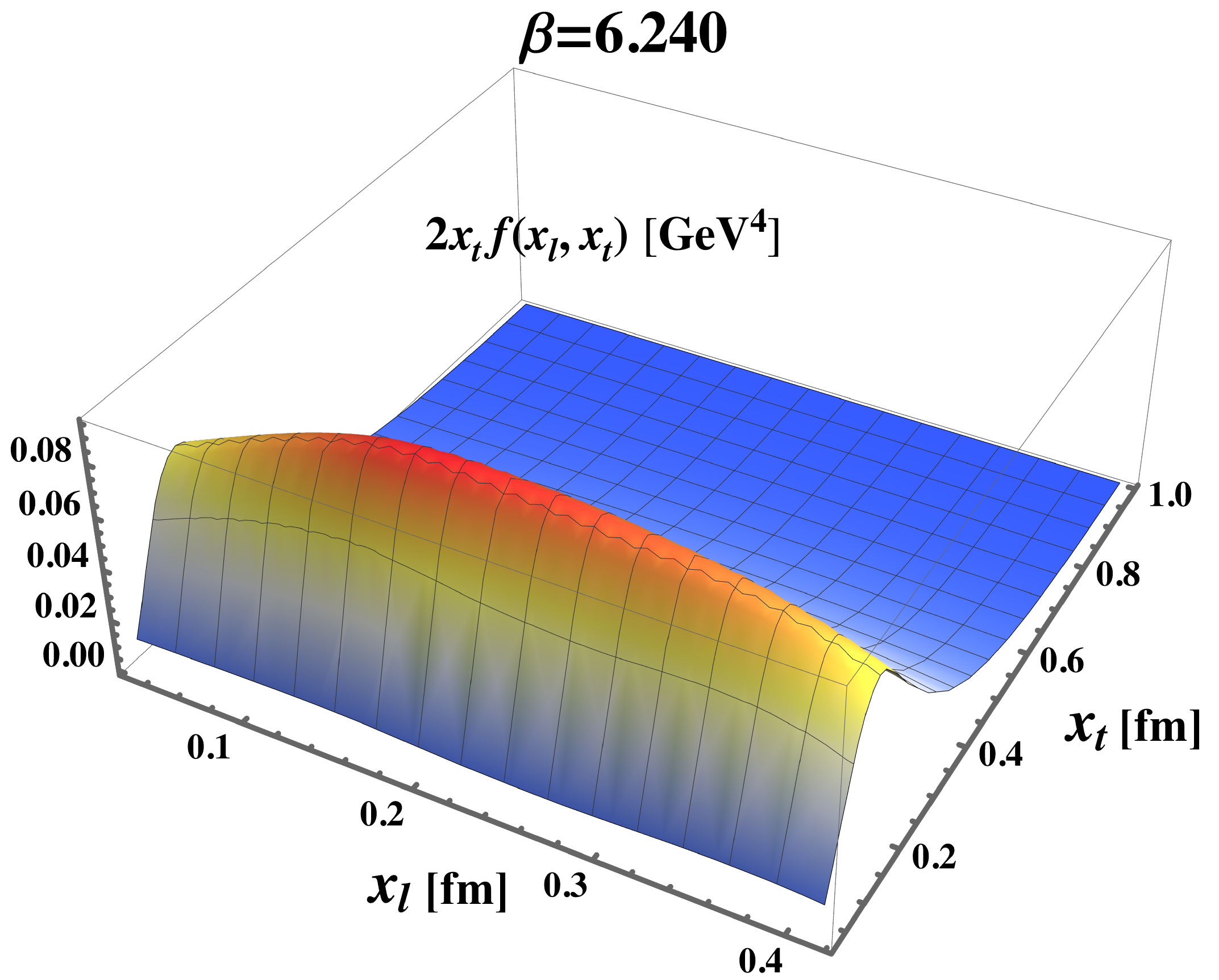}
\includegraphics[width=0.75\columnwidth,clip]{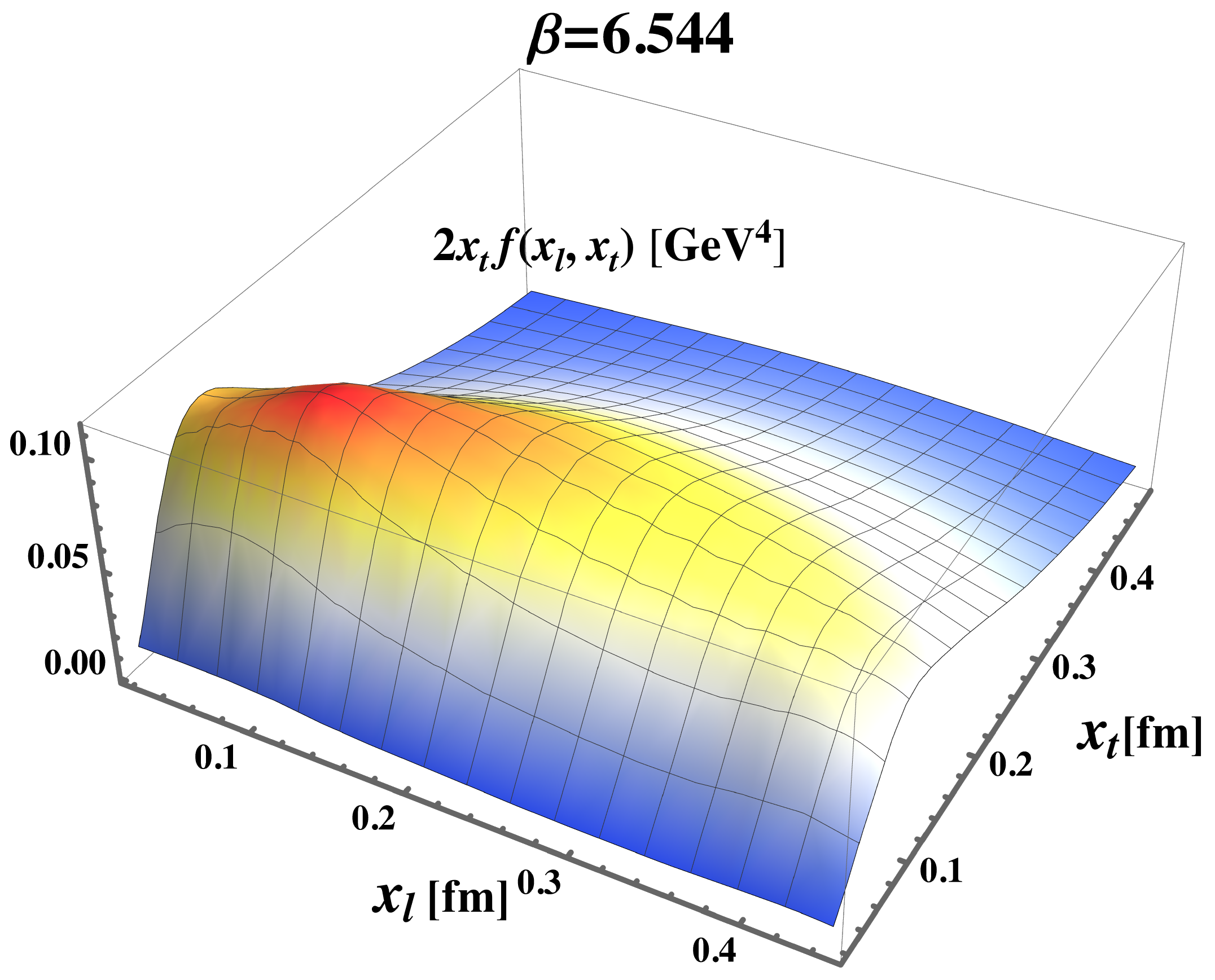}
\includegraphics[width=0.75\columnwidth,clip]{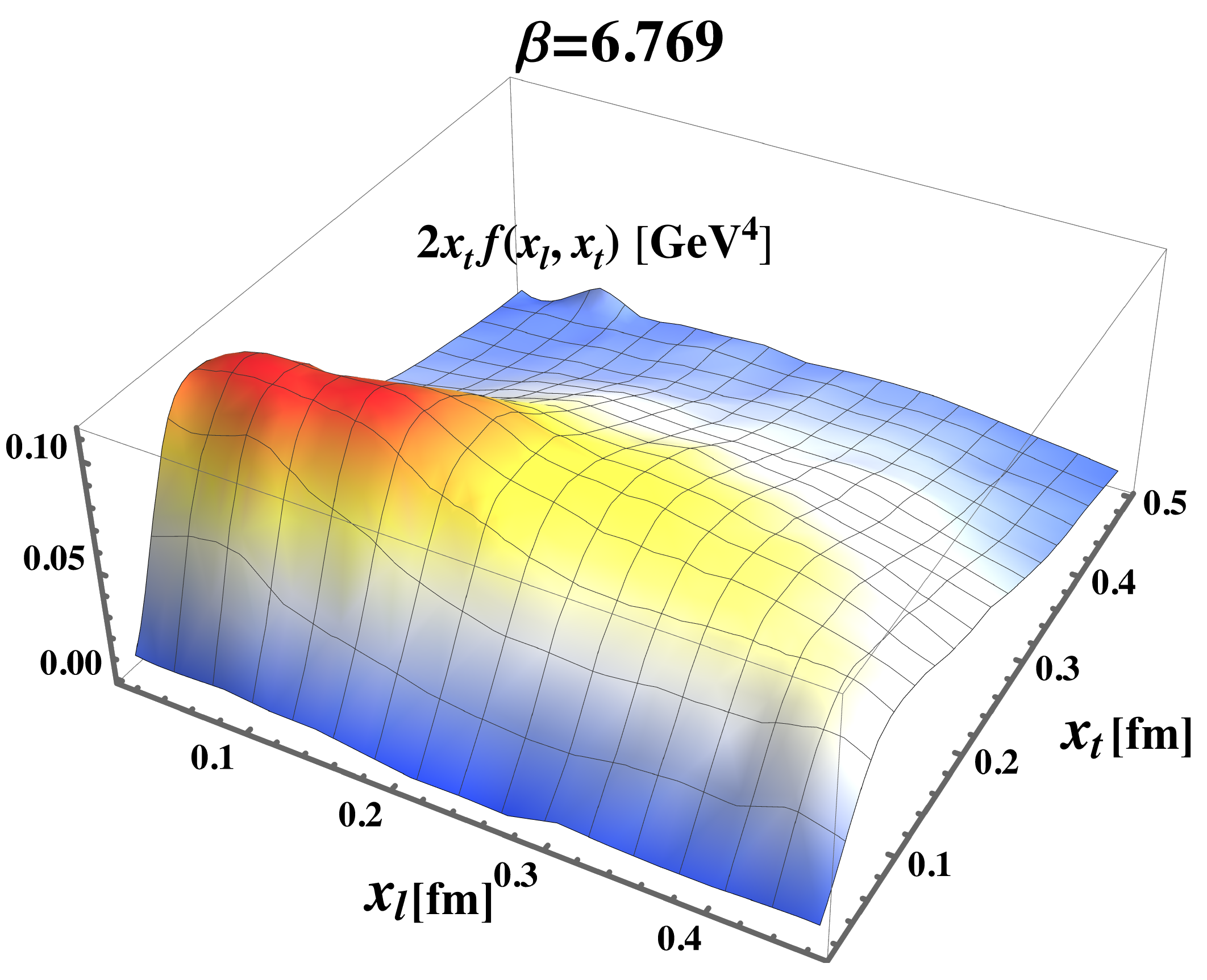}
\caption{The integrand of \Cref{vecF}, $2 x_t f(x_l, x_t)  = 2 x_t$ times the magnetic force density, for three values of $\beta$, all corresponding to the same quark-antiquark separation in physical units, $d =0.511 fm$.
The  force density falls off for large $x_t$ for all values of $x_l$, manifesting confinement. }
\label{fig:integrand}
\end{figure*}

Here we evaluate the magnitude of the force $\vec{F}$ and of the string tension $\sigma$  by numerically calculating the integrals in  \Cref{stress} and \Cref{vecF}.

First, we determine the integrand  of \Cref{vecF} over a grid of $(x_l,x_t)$ points, using the averages over ensembles of $J^{\rm mag}_z$ and $E_x^{{NP}}$. The behavior under smearing $J^{\rm mag}_z$
shows a fast growth up to a broad maximum, followed by a slow decay; under smearing $E_x^{(NP)}$
behaves similarly, except that the maximum is much less pronounced and is typically followed by an extended plateau. In both cases we chose as {\em optimal} smearing step as the one at the local maximum. Then, we spline-interpolate the discrete integrand by a smooth function using the tool {\tt Interpolation} of Wolfram Mathematica.

In \Cref{fig:integrand} we show the behavior of the integrand of \Cref{vecF} on the $(x_l,x_t)$-plane for three values of $\beta$. The numerical integration was finally performed using {\tt NIntegrate} in Mathematica.

Our results for  $\sqrt{F}$ are summarized in \Cref{force}. The statistical errors on the results were calculated by replicating the numerical integration 100 times after a flat reshuffling of the values of $J^{\rm mag}_z$ and $E_x^{\rm NP}$ entering the integrand of \Cref{vecF} within their respective 3$\sigma$ uncertainties, neglecting correlations between observables.

We also  estimated systematic errors on $\sqrt{F}$. Estimates of the positive systematic errors  were obtained by integrating over the variable $x_l$ up to the midpoint and then doubling the result assuming symmetry. This choice is motivated by our observation of an asymmetry, in the $x_l$ direction w.r.t. the midpoint, of the integrand of~\Cref{vecF} that is, in fact, systematically larger correspondingly to shorter Schwinger lines ($x_l<d/2$), as a result of an analogous asymmetry both in the magnetic current and in the non-perturbative field, which we ascribe to an interplay of renormalization and noise reduction through smearing.  Estimates of the negative systematic error (available only for the two larger values of $\beta$) were, instead, obtained by performing the integration in $x_l$ over the smallest domain, {\em i.e.} the region of integration in the case of $\beta=6.240$. By considering the smallest available domain, while clearly underestimating the actual result, we eliminate the systematic discrepancy between results at different lattice spacings coming from the fact that shorter and shorter distances from the sources can be attained on finer and finer lattices.

\Cref{force} also shows the calculation of $\sqrt{\sigma}$  from the non-perturbative field at the midpoint of the flux tube (See \Cref{stress}):
\begin{align}
&\sigma =  \int \mathrm{d}^2x_t 
  \frac{(E^{\mathrm{NP}}_x (x_l^{\mathrm{midpoint}},x_t))^2}{2}  \nonumber \\ &\approx \pi \int_0^{x_{t,\mathrm{max}}} \mathrm{d}x_t \, x_t \, (E^{\mathrm{NP}}_x (x_l^{\mathrm{midpoint}},x_t))^2 \;.
 \label{eq:stringtension}
 \end{align}
In \Cref{eq:stringtension}, after a spline interpolation of the numerical data for 
$E^{\mathrm{NP}}_x (x_l^{\mathrm{midpoint}},x_t))$, the integral has been numerically evaluated.
 
We observe that the resulting value for $\sqrt{F}$ remains fairly stable approaching the continuum and is only slightly larger than $\sqrt{\sigma}$, which is estimated to be
$0.464$~GeV~\cite{Necco:2001xg}. 

In the Appendix  we compare our results for the string tension with the value of the non-perturbative field $\vec{E}^{NP}$ at the position of the quark sources.

\section{Conclusions}
\label{sec:conclusion}

Using the connected correlator function $\rho_{W, \mu \nu}^{\rm conn} $ to provide a lattice definition of a gauge invariant field strength tensor $F_{\mu \nu}$ in SU(3) gauge theory, we have isolated
a non-perturbative gauge invariant longitudinal electric field, $\vec E^{NP}$, and magnetic currents,  $\vec J_{\rm mag}$, that circulate about the axis of the flux tube.  The magnetic current density $\vec{J}_{\rm mag} (x)$, in contrast to the electric charge density $\rho_{\rm el} (x)$, is manifest throughout the whole length of the flux tube.
The behavior of the expectation value of magnetic currents over smeared Monte Carlo ensembles suggests that they can be safely extrapolated to the continuum and have therefore the status of physical observables. 

The interaction of the magnetic currents with the non-perturbative longitudinal electric field gives rise to a magnetic Lorentz force density $\vec f=\vec J_{\rm mag} \times \vec E^{NP}~$filling the flux tube, directed toward the flux tube axis. \Cref{fig:integrand} shows the calculated magnitude of this  force density  rises to a maximum at $x_t \approx 0.1~ \rm{fm}$ all along the flux tube length, and then falls off for large $x_t$, manifesting confinement.

 Integrating the force density $\vec f$ over one half of a flux tube cut by a plane through its axis produces a force $\vec{F}$ on that half; integrating $\vec f$ over the other half of the tube produces an equal and opposite force on the other half, pushing the two halves together. Thus the flux tube is confined by these Maxwell-like forces on magnetic currents; the flux inside cannot spread out.
  
We find that the magnitude of the confining force $F$ is compatible, within systematic errors, to the magnitude  of the string tension $\sigma$, which is in turn determined by the distribution of the stress tensor on the  midplane perpendicular to the flux tube axis \Cref{eq:stringtension}, checking the consistency of our numerical calculations.

On the basis of our numerical simulations of field distributions in SU(3) gauge theory and using the analogy with the basic principles of electromagnetism, we have thus obtained a consistent physical picture of the interior of the flux tube, thereby unveiling confinement.

\section*{Acknowledgements}
This investigation was in part based on the MILC collaboration’s public lattice gauge theory code (\url{https://github.com/milc-qcd/}). Numerical calculations have been made possible through a CINECA-INFN agreement, providing access to HPC resources at CINECA. LC and AP acknowledge support from INFN/NPQCD project. FC and VC acknowledge support by the Deutsche Forschungsgemeinschaft (DFG, German Research Foundation) through the CRC-TR 211 ``Strong-interaction matter under extreme conditions'' -- project number 315477589 -- TRR 211. FC acknowledges the support by the State of Hesse within the Research Cluster ELEMENTS (Project ID 500/10.006).

\appendix

\section{The force and the field at the position of the source}
\label{sec:source}

We use \Cref{fig:op_W} and \Cref{connected1,connected2} of our paper to evaluate the field $E_x = F_{41}$ as follows.
Let $U_P = U_{41} (x_4=0,  x_1=x_l=a/2, x_t=0)$, {\em i.e.}, the center of the plaquette is the point $x_4 = 0$,  $x_1 = a/2$.
Then connect this plaquette to the center of the link $U_4^* (x_4=a/2,  x_1=0)$ of the Wilson loop $W$ in \Cref{fig:op_W} with the 
following choice of the Wilson Lines $L$ and $L^*$ in \Cref{fig:op_W}:
\begin{itemize}
    \item[$\bullet$] The Schwinger line $L$ in \Cref{fig:op_W} connecting the point $x_4=0$,  $x_1=0$ on $W$ to the point $x_4=a/2$,  $x_1=0$  on $U_{41}$;
    \item[$\bullet$] The Schwinger line $L^*$ in \Cref{fig:op_W} connecting the point $x_4=-a/2$ ,  $x_1=0$ on $U_{41}$ to the point $x_4 = 0$, on $W$.
\end{itemize}
With this choice of $U_P$,  $L$ and $L^*$, it then follows that
\begin{equation}
\tr (W L U_P L^*)  = \tr( W(C') )\;,
\label{eq:A1}
\end{equation}
where $C'$ is the loop bounding the area of a loop in which the area of the plaquette $U_P$ has been removed from the area of the original loop depicted in \Cref{fig:op_W}.

\Cref{eq:A1} follows from the fact that in the product $W L U_P$ the matrix $U^*$ connecting the point $x_4=a/2$  with the point $x_4=0$ in $W$  eliminates a corresponding matrix $U$ connecting these two points in $U_P$.  ($UU^* = \mathds{1}$).
Similarly  the product $U_P L^* W$ eliminates matrices connecting the points $x_4=-a/2$  and $x_4=0$  in $U_P$ and $W$.
As a consequence in the $\tr( W L U_P L)$ the lines connecting $x_4=a/2$ and $x_4=a/2$  in both $W$ and in $U_P$  have been eliminated.
There remain direct connections between $W$ and $U_P$ at $x_4=a/2$ and at $x_4=-a/2$,  yielding the result $\tr W(C')$, where $C'$ is a loop traversing the remaining three lines in $U_P$ and then connecting to $W$ at these points. This is result asserted in \Cref{eq:A1}.

Next note that to order $a^2$  the second term in \Cref{connected1} of our paper equals one since $\tr U_P=N$ to that order.
Thus if we multiply \Cref{connected1} by $\langle \tr(W)\rangle$ and use \Cref{eq:A1}, we obtain:
\begin{equation}
    \langle \tr(W) \rangle \ \rho^{\rm conn}_{{\rm W}, 41} = \langle \tr(W(C')) \rangle - \langle \tr(W) \rangle.
    \label{eq:A2}
\end{equation}
The left hand side of \Cref{eq:A2} determines the field at the position of the quark.
The right hand hand side of \Cref{eq:A2} determines the change in the heavy  quark potential when the piece of the Wilson loop between  $x_4=-a/2$ and $x_4=a/2$ is displaced by a distance $a$ in the $x$ direction.

Then, making use of \Cref{eq:A2}, we can calculate the derivative of the heavy quark potential, and, therefore, the string tension from the field 
at the position of the quark~\cite{Brambilla:2021wqs,Brambilla:2021egm},

\begin{equation}
    \sigma = - \frac{\partial_A  \left \langle \tr(W) \right \rangle}{\left \langle \tr(W) \right \rangle} = \frac{\rho^{\rm conn}_{{\rm W}, 41}}{a^2} = g E_x(0) \ , 
    \label{eq:A3}
\end{equation}
where $\partial_A$ denotes the derivative with respect to the area $A$.

This provides us with an alternative method to evaluate the string tension, giving the
following results:
\[
\begin{array}{ll}
\beta=6.240\;, & \;\;\;\;\; \sqrt{\sigma}=0.56353(81) {\rm \ GeV}\;, \\
\beta=6.544\;, & \;\;\;\;\; \sqrt{\sigma}=0.5962(38)  {\rm \ GeV}\;, \\
\beta=6.769\;, & \;\;\;\;\; \sqrt{\sigma}=0.617(16)   {\rm \ GeV}\;. \\
\end{array}
\]
These values are larger than our previous evaluations  (\Cref{force}) and do not exhibit scaling, reflecting the uncertainty in our calculations of $\vec{E}^{NP}$ close to the sources.

\bibliography{qcd}

\begin{thebibliography}{44}

\bibitem{greensite2011introduction}
J.~Greensite, \emph{{An introduction to the confinement problem}}, Vol. 821
  (Springer Berlin Heidelberg, 2011)

\bibitem{Diakonov:2009jq}
D.~Diakonov, Nucl. Phys. Proc. Suppl. \textbf{195}, 5 (2009),
  \texttt{0906.2456}

\bibitem{Bander:1980mu}
M.~Bander, Phys. Rept. \textbf{75}, 205 (1981)

\bibitem{Greensite:2003bk}
J.~Greensite, Prog. Part. Nucl. Phys. \textbf{51}, 1 (2003),
  \texttt{hep-lat/0301023}

\bibitem{Ripka:2005cr}
G.~Ripka, AIP Conf. Proc. \textbf{775}, 262 (2005)

\bibitem{Simonov:2018cbk}
Y.A. Simonov (2018), \texttt{1804.08946}

\bibitem{Fukugita:1983du}
M.~Fukugita, T.~Niuya, Phys. Lett. \textbf{B132}, 374 (1983)

\bibitem{Kiskis:1984ru}
J.E. Kiskis, K.~Sparks, Phys. Rev. \textbf{D30}, 1326 (1984)

\bibitem{Flower:1985gs}
J.W. Flower, S.W. Otto, Phys. Lett. \textbf{B160}, 128 (1985)

\bibitem{Wosiek:1987kx}
J.~Wosiek, R.W. Haymaker, Phys. Rev. \textbf{D36}, 3297 (1987)

\bibitem{DiGiacomo:1989yp}
A.~Di~Giacomo, M.~Maggiore, S.~Olejnik, Phys. Lett. \textbf{B236}, 199 (1990)

\bibitem{DiGiacomo:1990hc}
A.~Di~Giacomo, M.~Maggiore, S.~Olejnik, Nucl. Phys. \textbf{B347}, 441 (1990)

\bibitem{Cea:1992sd}
P.~Cea, L.~Cosmai, Nucl. Phys. Proc. Suppl. \textbf{30}, 572 (1993)

\bibitem{Matsubara:1993nq}
Y.~Matsubara, S.~Ejiri, T.~Suzuki, Nucl. Phys. Proc. Suppl. \textbf{34}, 176
  (1994), \texttt{hep-lat/9311061}

\bibitem{Cea:1994ed}
P.~Cea, L.~Cosmai, Phys. Lett. \textbf{B349}, 343 (1995),
  \texttt{hep-lat/9404017}

\bibitem{Cea:1995zt}
P.~Cea, L.~Cosmai, Phys. Rev. \textbf{D52}, 5152 (1995),
  \texttt{hep-lat/9504008}

\bibitem{Bali:1994de}
G.S. Bali, K.~Schilling, C.~Schlichter, Phys. Rev. \textbf{D51}, 5165 (1995),
  \texttt{hep-lat/9409005}

\bibitem{Green:1996be}
A.M. Green, C.~Michael, P.S. Spencer, Phys. Rev. \textbf{D55}, 1216 (1997),
  \texttt{hep-lat/9610011}

\bibitem{Skala:1996ar}
P.~Skala, M.~Faber, M.~Zach, Nucl.Phys. \textbf{B494}, 293 (1997),
  \texttt{hep-lat/9603009}

\bibitem{Haymaker:2005py}
R.W. Haymaker, T.~Matsuki, Phys. Rev. \textbf{D75}, 014501 (2007),
  \texttt{hep-lat/0505019}

\bibitem{D'Alessandro:2006ug}
A.~D'Alessandro, M.~D'Elia, L.~Tagliacozzo, Nucl.Phys. \textbf{B774}, 168
  (2007), \texttt{hep-lat/0607014}

\bibitem{Cardaci:2010tb}
M.S. Cardaci, P.~Cea, L.~Cosmai, R.~Falcone, A.~Papa, Phys.Rev. \textbf{D83},
  014502 (2011), \texttt{1011.5803}

\bibitem{Cea:2012qw}
P.~Cea, L.~Cosmai, A.~Papa, Phys.Rev. \textbf{D86}, 054501 (2012),
  \texttt{1208.1362}

\bibitem{Cea:2013oba}
P.~Cea, L.~Cosmai, F.~Cuteri, A.~Papa, PoS \textbf{LATTICE2013}, 468 (2013),
  \texttt{1310.8423}

\bibitem{Cea:2014uja}
P.~Cea, L.~Cosmai, F.~Cuteri, A.~Papa, Phys. Rev. \textbf{D89}, 094505 (2014),
  \texttt{1404.1172}

\bibitem{Cea:2014hma}
P.~Cea, L.~Cosmai, F.~Cuteri, A.~Papa, PoS \textbf{LATTICE2014}, 350 (2014),
  \texttt{1410.4394}

\bibitem{Cardoso:2013lla}
N.~Cardoso, M.~Cardoso, P.~Bicudo, Phys. Rev. \textbf{D88}, 054504 (2013),
  \texttt{1302.3633}

\bibitem{Caselle:2014eka}
M.~Caselle, M.~Panero, R.~Pellegrini, D.~Vadacchino, JHEP \textbf{01}, 105
  (2015), \texttt{1406.5127}

\bibitem{Cea:2015wjd}
P.~Cea, L.~Cosmai, F.~Cuteri, A.~Papa, JHEP \textbf{06}, 033 (2016),
  \texttt{1511.01783}

\bibitem{Cea:2017ocq}
P.~Cea, L.~Cosmai, F.~Cuteri, A.~Papa, Phys. Rev. \textbf{D95}, 114511 (2017),
  \texttt{1702.06437}

\bibitem{Shuryak:2018ytg}
E.~Shuryak (2018), \texttt{1806.10487}

\bibitem{Bonati:2018uwh}
C.~Bonati, S.~Cal\`i, M.~D'Elia, M.~Mesiti, F.~Negro, A.~Rucci, F.~Sanfilippo,
  Phys. Rev. \textbf{D98}, 054501 (2018), \texttt{1807.01673}

\bibitem{Shibata:2019bke}
A.~Shibata, K.I. Kondo, S.~Kato (2019), \texttt{1911.00898}

\bibitem{Baker:2018mhw}
M.~Baker, P.~Cea, V.~Chelnokov, L.~Cosmai, F.~Cuteri, A.~Papa, Eur. Phys. J.
  \textbf{C79}, 478 (2019), \texttt{1810.07133}

\bibitem{Baker:2019gsi}
M.~Baker, P.~Cea, V.~Chelnokov, L.~Cosmai, F.~Cuteri, A.~Papa, Eur. Phys. J. C
  \textbf{80}, 514 (2020), \texttt{1912.04739}

\bibitem{Yanagihara:2018qqg}
R.~Yanagihara, T.~Iritani, M.~Kitazawa, M.~Asakawa, T.~Hatsuda, Phys. Lett.
  \textbf{B789}, 210 (2019), \texttt{1803.05656}

\bibitem{Yanagihara:2019foh}
R.~Yanagihara, M.~Kitazawa, PTEP \textbf{2019}, 093B02 (2019),
  \texttt{1905.10056}

\bibitem{Bikudo:2018}
P.~Bicudo, N.~Cardoso, M.~Cardoso, Phys. Rev. D \textbf{98}, 114507 (2018)

\bibitem{Mueller:2019mkh}
L.~Mueller, O.~Philipsen, C.~Reisinger, M.~Wagner (2019), \texttt{1907.01482}

\bibitem{Necco:2001xg}
S.~Necco, R.~Sommer, Nucl. Phys. \textbf{B622}, 328 (2002),
  \texttt{hep-lat/0108008}

\bibitem{Hasenfratz:2001hp}
A.~Hasenfratz, F.~Knechtli, Phys. Rev. \textbf{D64}, 034504 (2001),
  \texttt{hep-lat/0103029}

\bibitem{Battelli:2019lkz}
N.~Battelli, C.~Bonati, Phys. Rev. \textbf{D99}, 114501 (2019),
  \texttt{1903.10463}

\bibitem{Brambilla:2021wqs}
N.~Brambilla, V.~Leino, O.~Philipsen, C.~Reisinger, A.~Vairo, M.~Wagner, Phys.
  Rev. D \textbf{105}, 054514 (2022), \texttt{2106.01794}

\bibitem{Brambilla:2021egm}
N.~Brambilla, H.S. Chung, A.~Vairo, X.P. Wang, JHEP \textbf{01}, 184 (2022),
  \texttt{2111.07811}

\end{thebibliography}

\end{document}